\documentclass{vgtc}                          % final (conference style)
\graphicspath{{figures/}{pictures/}{images/}{./}} % where to search for the images

\usepackage{times}                     % we use Times as the main font
\usepackage{tabu}                      % only used for the table example
\usepackage{booktabs}                  % only used for the table example
\usepackage{lipsum}                    % used to generate placeholder text
\usepackage{mwe}                       % used to generate placeholder figures

\usepackage{mathptmx}                  % use matching math font
\usepackage{CJKutf8}

\onlineid{7862}

\vgtccategory{Research}

\vgtcinsertpkg

\usepackage{xspace}
\newcommand{\ours}{\emph{RAGE-Vis}\xspace}

\title{\ours: A Relation-Aware Generative Editing Interface for Natural Language-Based Chart Editing}

\author{Ziyao Kang, Yiping Sun, Linxuan Tian, Henghuan Qu, Wei Zeng, and Jiazhi Xia*}

\authorfooter{%
\textbullet\ Ziyao Kang, Yiping Sun, Linxuan Tian, Henghuan Qu, and Jiazhi Xia are with the Central South University. E-mail: \{kangzy, yipingsun, tianlinxuan, 8203240620, xiajiazhi\}@csu.edu.cn.\\
\textbullet\ Wei Zeng is with the Hong Kong University of Science and Technology (Guangzhou) and the Hong Kong University of Science and Technology. E-mail: weizeng@hkust-gz.edu.cn.\\
\textbullet\ Jiazhi Xia is the corresponding author.
}

\makeatletter
\renewcommand{\copyrightspace}{%
  \renewcommand{\thefootnote}{}%
  \begingroup
  \renewcommand{\@makefntext}[1]{\noindent ##1}%
  \footnotetext[0]{\authorfootertext}%
  \endgroup
  \renewcommand{\thefootnote}{\arabic{footnote}}%
}
\makeatother

\teaser{
  \centering
  \includegraphics[width=0.95\linewidth]{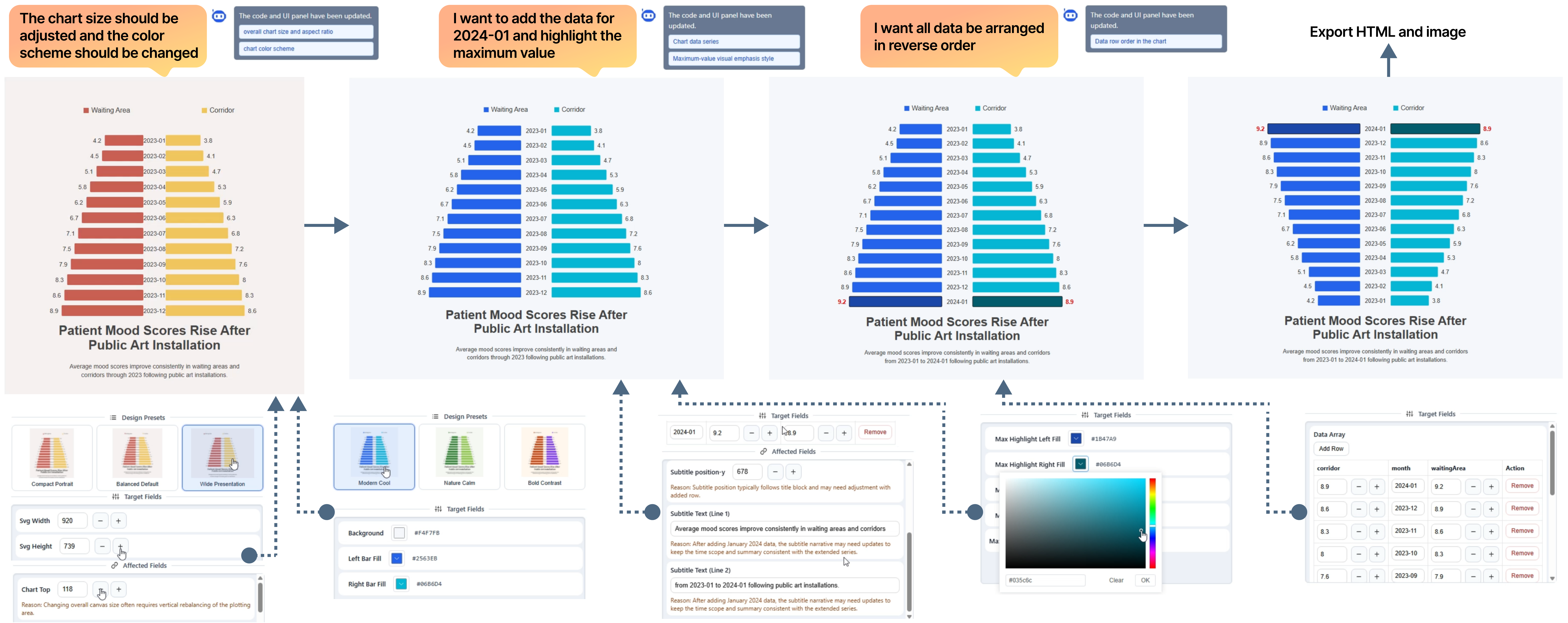}
  \caption{\textbf{Case I: \ours supports relation-aware editing of a chart image.} Starting from an uploaded mirrored bar chart, the system parses the visual content, reconstructs it into editable code, and generates a visual editing panel. Users can then iteratively issue natural-language requests, such as resizing the chart and changing its color style, adding new data and highlighting the maximum value, and reversing the data order. For each request, the system exposes both target fields and affected fields, enabling coordinated controls across related components.} 
  \label{fig:teaser}
}

\abstract{
    Natural language offers an easy way for users to express chart editing intents, which are often composite and cross‑component (e.g., adjusting style, extending categories, highlighting values). 
    However, existing methods typically map instructions to a single operation or widget, limiting their ability to handle high‑level requests and often producing locally plausible but globally inconsistent results due to a lack of awareness of relationships between chart components.
    To address these challenges, we introduce \ours, a \emph{R}elation-\emph{A}ware \emph{G}enerative \emph{E}diting interface for natural language-based chart editing. 
    The system supports bitmap chart images 
    as input and converts them into an editable parameterized intermediate representation. 
    Instead of mapping instructions to a single edit or widget, \ours parses composite intents, identifies targets and scopes, and generates hierarchical editing panels for underspecified requests, enabling users to adjust both global settings and local parameters.
    Furthermore, \ours identifies potentially affected fields based on visual encoding relations, structural relations, and expressive consistency relations, and organizes them into actionable widgets to support cross-component coordinated controls. 
    Through two case studies, we demonstrate the applicability of \ours in complex editing tasks, including style adjustment, data extension, order rearrangement, legend layout, and color mapping.
    A user study further shows that participants can effectively handle underspecified requests, explore candidate alternatives, and maintain cross-component consistency with \ours.

} % end of abstract


\begin{document}


\firstsection{Introduction}

\maketitle

%% \section{Introduction} %for journal use above \firstsection{..} instead

%图表是数据分析和信息传播中最普遍的可视化形式之一。可视化创作工具的进步使用户能够通过声明式语法、直接操作界面或自动生成流程来构建和优化图表。然而，实际的编辑需求往往表现为包含高层次目标、描述不够具体以及多个子任务的复合请求。例如，“使图表更协调”、“突出显示峰值”或“在保持风格一致性的前提下添加新类别”。虽然这些自然语言指令对用户来说很直观，但在解码用户意图、准确识别目标对象以及保持图表各组件之间的一致性方面，却带来了技术挑战。
Charts are among the most pervasive visualization forms for data analysis and information dissemination. 
Advances in visualization authoring tools enable users to construct and refine charts via declarative grammars, direct manipulation interfaces, or automated generation pipelines \cite{7539624, 10.1145/3411764.3445356, DBLP:journals/cgf/SatyanarayanH14, 8440827, 10.1145/3173574.3173697, DBLP:journals/corr/abs-2209-08834, 10.1145/3613904.3642639, 11262789}. 
% Nevertheless, practical editing requirements typically transcend isolated, explicit parameter modifications. 
However, practical editing needs often appear as composite requests with high-level objectives, underspecified descriptions, and multiple subtasks.
For example, ``make the chart more coordinated'', ``highlight the peak value'', or ``add a new category while maintaining stylistic consistency''. 
While intuitive for users, such natural language instructions pose technical challenges in decoding user intent, accurately identifying target objects, and maintaining consistency across interconnected chart components.

%越来越多的研究探索了用于可视化的自然语言界面（NLI），结果表明自然语言能够显著降低表达图表创作和编辑意图的门槛。然而，纯粹基于语言的交互往往会牺牲图形用户界面（GUI）所提供的精细控制、即时视觉反馈和可重复性。为了弥补这一差距，以往的系统引入了动态生成的用户界面组件或混合输入交互机制。DynaVis进一步展示了这种混合范式的价值：通过动态合成持久的用户界面组件，它允许用户迭代地调整参数并获得即时反馈，从而减轻了用户回忆精确命令或导航复杂菜单的认知负担。更广泛地说，PUMICE以及基于草图的视觉查询研究表明，结合多种交互机制有助于缩小高级目标与可执行操作之间的差距。总而言之，这些研究表明，将自然语言的表达能力与可直接操作组件的可控性相结合，是自然语言驱动的图表编辑的一个很有前景的方向。
A growing body of research 
% indicates that neither Natural Language Interfaces (NLIs) nor traditional Graphical User Interfaces (GUIs) alone are sufficient for efficient visualization editing; 
% instead, users require a synergy of both. 
has explored natural language interfaces (NLIs) for visualization \cite{10.1145/2984511.2984588, 9912366}, demonstrating that natural language can significantly lower the barrier for expressing chart authoring and editing intents. 
However, purely language-based interaction often sacrifices the fine-grained control, immediate visual feedback, and repeatability afforded by GUIs.
To bridge this gap, 
% systems like DataTone \cite{10.1145/2807442.2807478}, NL2Interface \cite{DBLP:journals/corr/abs-2209-08834}, and the Data Formulator series \cite{10292609, 10.1145/3706598.3713296} 
prior systems have introduced dynamically generated UI widgets or mixed-input interaction mechanisms \cite{10.1145/2807442.2807478, DBLP:journals/corr/abs-2209-08834, 10292609, 10.1145/3706598.3713296, liu2017towards}. 
% These approaches allow users to specify initial intents through natural language and subsequently utilize direct manipulation interfaces to fine-tune parameters and refine outputs. 
DynaVis \cite{10.1145/3613904.3642639} further demonstrates the value of this hybrid paradigm: by dynamically synthesizing persistent UI widgets, it allows users to iteratively adjust parameters and receive immediate feedback, reducing the cognitive burden of recalling precise commands or navigating complex menus. 
More broadly, PUMICE \cite{10.1145/3332165.3347899} and research on sketch-based visual queries \cite{8807280} suggest that combining multiple interaction mechanisms helps narrow the gulf of execution between high-level goals and executable operations. 
% In summary, the core user need is not a specific interaction modality but rather the ability to explore design alternatives, inspect feedback, and iterate rapidly with minimal friction. 
Together, these works indicate that combining the expressiveness of natural language with the controllability of directly manipulable widgets is a promising direction for natural language-driven chart editing.

%然而，仅仅将自然语言请求翻译成可操作的控件不足以应对复杂的图表编辑。图表并非一系列孤立的视觉元素的集合。其组成部分，包括标记、图例、坐标轴、标签和注释，通过视觉编码、结构依赖关系和表达目标相互关联。因此，编辑一个组件可能需要对相关组件进行协调控制。如果在编辑过程中未能保留这些关联，即使系统成功生成了交互式控件，其结果也可能局部有效但全局不一致。
However, simply translating natural language requests into actionable widgets is insufficient for complex chart editing.
A chart is not a disparate collection of isolated visual elements. Its components, including marks, legends, axes, labels, and annotations, are connected through visual encodings, structural dependencies, and expressive goals. 
As a result, editing one component may require coordinated controls to related components. 
Without preserving such relationships during the editing process, a system may yield results that are locally valid but globally inconsistent, even if it successfully generates interactive widgets. 
% To address this, this paper explores the integration of component relationships into generative editing interfaces, enabling natural language intents, candidate widgets, and coordinated controls to operate synergistically within a unified chart semantic framework.

%为解决上述问题，本文提出 ChartUI，一个面向自然语言图表编辑的关系感知生成式界面。ChartUI 的核心思想是：自然语言图表编辑不应仅被视为指令到控件的映射，而应被表示为一个由用户意图、图表语义和组件关系共同驱动的交互式编辑过程。系统首先解析用户输入中的复合编辑意图，将其拆解为多个子意图，并识别其中未明确指定的目标对象、参数和作用范围。在此基础上，ChartUI 针对模糊或欠指定请求生成层级式候选编辑面板，使用户能够比较不同方案，并在全局策略基础上继续进行局部修正。进一步地，系统根据图表组件之间的关系，识别视觉编码一致性、结构组成和表达一致性关系，并据此生成联动更新策略，以保证编辑结果在数据、视觉编码和表达上的一致性。
To address these challenges, we introduce \ours, a Relation-Aware Generative Editing interface for natural language-based chart editing. 
% The central premise of \ours is that natural language chart editing should be conceptualized not merely as a direct mapping from instructions to widgets, but as an interactive process driven by the synergy of user intent, chart semantics, and component relationships. 
\ours first parses composite editing intents from user input, decomposes them into multiple sub-intents, and identifies underspecified target objects, parameters, and operational scopes.
Based on the parsed intents, the system {\color{black}generates global-to-local editing panels} for underspecified requests, enabling users to compare diverse design alternatives and refine local parameters after selecting global strategies. 
Furthermore, \ours identifies relationships among chart components regarding visual encoding consistency, structural integrity, and expressive consistency. 
By leveraging these dependencies, the system enables coordinated controls across related components to ensure globally coherent editing outcomes.
We evaluate \ours through two case studies and a within-subjects user study. The case studies demonstrate its applicability to relation-aware editing across mirrored bar and radar charts. The user study compares \ours with a baseline without Design Presets and Affected Fields, showing that \ours reduces additional natural language requests and model response time, while better supporting alternative comparison, related-component discovery, and consistency maintenance.

%本文的主要贡献如下：
%1. 我们设计并实现ChartUI，一个面向自然语言图表编辑的关系感知生成界面系统，将复合意图解析、候选方案支持和跨组件联动编辑整合到统一流程中。 
%2. 我们构建一种图表组件关系方法，用于显式表示视觉编码关系、结构组成关系和表达一致性关系，用于支持复杂编辑中的一致性维护。 
%3.  我们通过case study和user study评估ChartUI在处理欠指定编辑请求、生成可操作候选方案以及维护跨组件一致性方面的effectiveness and usability。
The main contributions of this work are:
\begin{itemize}
  \item 
    We design and implement \ours that integrates composite intent parsing, candidate panel generation, and cross-component coordinated editing into a unified workflow.
  \item 
    We propose a chart-component relation approach that captures visual encoding, structural, and expressive consistency relations to support cross-component controls.
  \item 
    We evaluate the effectiveness of \ours through two case studies and a user study, demonstrating its ability to handle underspecified editing requests, generate actionable alternatives, and maintain cross-component consistency.
\end{itemize}

\section{Related Work}

\subsection{Chart Editing}
%图表编辑旨在支持用户对已有可视化的数据、视觉编码、布局、文本标注、样式进行调整。
Chart editing aims to support users in modifying the data, visual encodings, layouts, textual annotations, and styles of existing visualizations. 
%早期工作主要通过声明式语法与模板复用[40,41]、直接操作式图表构建[42-44]、D3 反解析与样式重设[45]、示例驱动合成[46]等方式提升图表可编辑性。尽管这些工作赋予了图表较高的结构可塑性，但通常要求用户明确指定编辑对象、视觉通道或示例映射，难以自然承接非专家用户的高层模糊意图。
Early work improved chart editability through declarative grammars and template reuse \cite{7539624, 10.1145/3411764.3445356}, direct-manipulation chart construction \cite{DBLP:journals/cgf/SatyanarayanH14, 8440827, 10.1145/3173574.3173697}, D3 deconstruction and restyling \cite{10.1145/2642918.2647411}, and example-driven synthesis \cite{10.1145/3411764.3445249}. 
Although these approaches provide substantial structural plasticity, they typically require users to explicitly specify editing targets, visual channels, or example mappings, falling short in accommodating high-level and underspecified intentions from non-expert users.
%随着交互技术与人工智能的发展，现有工作经历了从底层约束操作到高层意图驱动的演进。
With the development of interaction techniques and AI, existing work has evolved from low-level constrained operations toward high-level intent-driven paradigms.
%为进一步降低门槛，研究者开始利用自然语言接口与 AI 代理捕获编辑意图。相关工作已覆盖动态界面生成[47]、高层意图驱动的数据变换[15,18]，以及配色、金融叙事标注、位图图表重设计和静态图表动态化等具体编辑任务[48-51]。然而，代表性意图驱动系统仍存在灵活性限制：DynaVis[52]可根据自然语言指令合成可持续交互的 UI 控件，支持用户继续微调参数，但尚未显式将模糊或复合意图拆解为子意图；DataWink[19]支持基于示例的样式复用，但效果依赖参考图表质量，在缺乏明确视觉参照时，对纯文本驱动的欠指定意图支持仍然有限。这些系统多倾向于将复合指令整体映射，未能提供一种显式机制来分层展示模糊意图下的多个候选方案。
% To further lower the barrier to entry, researchers have begun to leverage natural language interfaces and AI agents to capture editing intentions. 
Existing work covers dynamic interface generation \cite{DBLP:journals/corr/abs-2209-08834}, high-level intent-driven data transformation \cite{10292609, 10.1145/3706598.3713296}, and specific editing tasks such as financial narrative annotation, bitmap chart redesign, and static-to-live chart transformation \cite{10292693, 10787087, 10.1145/2047196.2047247, 10530507}.

However, representative intent-driven systems still face limitations in flexibility. 
For example, DynaVis \cite{10.1145/3613904.3642639} synthesizes persistent interactive UI widgets from natural language instructions and allows users to further refine parameters.
However, it does not explicitly decompose vague or composite intents into sub-intents. 
DataWink \cite{11262789} supports example-based style reuse, but its effectiveness depends on the quality of reference charts and remains limited in supporting text-driven underspecified intentions when explicit visual references are absent. 
These systems tend to map composite instructions as a whole, lacking an explicit mechanism to hierarchically present candidate solutions for underspecified intents.

%近年来，多模态大模型被用于探索图表编辑任务，但 ChartM3[53]、FigEdit[54]、ChartEdit[55]与 ChartEditBench[56]等评测表明，图表是受图形语法约束的结构化表达，模型在语义一致性、结构保持、视觉定位和多轮状态维护上仍易出错。尽管 Bonás 等[57]提出了结合图表反解析、设计推理和迭代修复的框架，以及 MisVisFix[58]尝试通过反解析、诊断和交互式修正支持图表质量改进，但复杂编辑中的跨组件影响识别与一致性维护仍待进一步探索。
Recently, MLLMs have been used to explore chart editing tasks \cite{zeng2024advancing}. 
Benchmarks such as ChartM3 \cite{10.1145/3746027.3755714}, FigEdit \cite{DBLP:journals/corr/abs-2512-00752}, ChartEdit \cite{zhao-etal-2025-chartedit}, and ChartEditBench \cite{DBLP:journals/corr/abs-2602-15758} show that charts are structured expressions constrained by graphical grammars.
Models remain prone to errors in semantic consistency, structural preservation, visual localization, and multi-turn state maintenance. 
% Although Bonás et al. \cite{DBLP:journals/corr/abs-2602-20291} proposed a framework combining chart de-rendering, design reasoning, and iterative repair, and 
% MisVisFix \cite{11269882} attempts to support chart quality improvement through de-rendering, diagnosis, and interactive correction, cross-component impact identification and consistency maintenance in complex editing remain underexplored.
%综上，现有工作在执行明确编辑意图方面已取得进展，但在模糊意图的分层解析与跨组件一致性维护方面仍面临挑战。本文关注复合指令拆解与关系感知联动编辑，以在保障全局一致性的同时支持精准高效的图表编辑。
In summary, existing work has made progress in executing explicit editing intentions, but challenges remain in the hierarchical parsing of ambiguous intentions and the maintenance of cross-component consistency. 
Our work focuses on decomposing composite instructions and enabling relation-aware coordinated editing to support precise and efficient chart editing while preserving global consistency.

\subsection{Natural Language Interfaces for Visualization}
%自然语言接口已成为数据可视化中重要的交互范式之一[11]。早期工作主要关注将自然语言映射为数据属性、分析任务和可视化规格[1,2,8]。其中，DataTone 通过 ambiguity widgets 暴露解析与生成过程中的歧义，并支持交互式修正[2]；NL4DV 则将自然语言查询转化为包含数据属性、分析任务和 Vega-Lite 规格的结构化表示[8]。随后，ncNet、Chat2VIS、LIDA 与 ChartGPT 借助神经机器翻译或大语言模型，进一步提升了自然语言驱动的可视化生成能力[9,13,14,17]。这些工作奠定了自然语言可视化的基础，但总体上仍以单轮查询理解和可视化生成为主要目标。
Natural language interfaces (NLIs) have become an important interaction paradigm in data visualization \cite{9699035}. 
Early work mainly focused on mapping natural language to data attributes, analytical tasks, and visualization specifications \cite{10.1007/978-3-642-13544-6_18, 10.1145/2807442.2807478, 9222342}. 
% DataTone exposes ambiguities in parsing and generation through ambiguity widgets and supports interactive correction \cite{10.1145/2807442.2807478}; NL4DV transforms natural language queries into structured representations containing data attributes, analytical tasks, and Vega-Lite specifications \cite{9222342}. 
% Subsequently, ncNet, Chat2VIS, LIDA, and ChartGPT further 
Later works improved natural language-driven visualization generation with neural machine translation models or LLMs \cite{9617561, 10121440, dibia-2023-lida, 10443572}. 
These works laid the foundation for natural language interaction in visualization, but they still primarily target single-turn query understanding and visualization generation.
%随后，研究逐渐从单轮生成扩展到基于已有视图和分析上下文的连续交互，使用户能够结合当前可视化状态、历史查询或交互上下文持续探索数据[3-7,10]。其中，Setlur 等通过 inferencing 机制补全聚合、分组、过滤和排序等隐含信息[6]；FlowSense 将自然语言接口嵌入 dataflow visualization system，使用户能够结合数据流图上下文构建和调整分析流程[7]。这类工作推动自然语言接口从单轮查询走向持续交互，但重点仍放在分析查询的上下文理解与流程组织上。
Recent efforts have gradually expanded from single-turn interaction to sustained interaction based on existing views and analytical contexts, enabling users to continuously explore data by referring to the current visualization state, query history, or interaction context \cite{10.1145/2984511.2984588, 10.1145/3025171.3025227, 8019860, 10.1145/3301275.3302270, 8807265, 9973222, LI2025100241}. 
% Setlur et al. \cite{10.1145/3301275.3302270} use inferencing to complete implicit information such as aggregation, grouping, filtering, and sorting, while FlowSense \cite{8807265} embeds a natural language interface into a dataflow visualization system so that users can construct and adjust analytical workflows in the context of dataflow graphs. 
While these works transition NLIs from single-turn queries toward sustained interaction, their primary focus remains on the contextual understanding of user queries and organization of analytical workflows.

%近年的研究进一步将自然语言用于可视化创作与编辑。Wang 等人将可视化编辑意图形式化为可执行的 editing actions，推动研究从“自然语言查询”走向“自然语言编辑”[12]。Data Formulator 系列与 DataWink 分别从混合输入式创作和 SVG 示例复用角度，探索自然语言或多模态大模型辅助的可视化创作[15,18,19]。此外，VisEval 等 benchmark 的出现也表明，自然语言驱动的可视化生成已进入系统化评测阶段[16]。
Recent studies have further applied natural language to visualization authoring and editing \cite{lu2025large}. 
Wang et al. \cite{9912366} formalize visualization editing intents as executable editing actions, shifting the focus from natural language querying to natural language editing. 
The Data Formulator series explores mixed-input visualization authoring with natural language support, while DataWink leverages large multimodal models to support SVG example reuse \cite{10292609, 10.1145/3706598.3713296, 11262789}.
The emergence of benchmarks such as VisEval indicates that natural language-driven visualization generation has entered a stage of systematic evaluation \cite{10670425}.
%现有研究已能够将自然语言用于可视化生成、分析 refinement 与创作辅助，但多数仍主要处理相对明确、边界清晰的单一用户意图，且较少考虑相关图表组件之间的联动影响。本文面向更复杂的自然语言图表编辑场景，关注如何将复合或欠指定语言输入转化为可操作的分层编辑过程，并结合图表组件关系支持跨组件联动控制。
% Existing research has enabled natural language to support visualization generation, analytical refinement, and authoring assistance, but 
However, most existing work primarily addresses relatively explicit and well-bounded user intents, largely overlooking the necessary coordination across interdependent chart components. 
Our work targets more complex natural language-driven chart editing scenarios. 
It transforms composite or underspecified language inputs into actionable hierarchical editing processes and supports cross-component coordinated controls based on chart component relations.

\subsection{Generative User Interfaces}
%早期界面生成研究主要通过任务和领域模型生成可执行界面[20-23,25]。MIKE[20]、ITS[21]、HUMANOID[22]和 MASTERMIND[23]等系统分别从语义定义、逐步细化、多层抽象、原型生成与迭代设计等方面探索了模型驱动的界面生成；Puerta 等人[25]指出，抽象模型到具体界面元素的映射仍是关键瓶颈。这类方法有助于维护开发一致性，但依赖预先建模，门槛较高，且模型相对静态。
Early interface generation research mainly generated executable interfaces from task and domain models \cite{10.1145/27623.28868, 10.1145/98188.98194, 10.1145/142750.142912, Szekely1996, DBLP:conf/iui/PuertaE99}. 
% Systems such as MIKE \cite{10.1145/27623.28868}, ITS \cite{10.1145/98188.98194}, HUMANOID \cite{10.1145/142750.142912}, and MASTERMIND \cite{Szekely1996} explored model-driven interface generation from the perspectives of semantic definitions, stepwise refinement, multi-level abstraction, prototype generation, and iterative design. 
% Puerta et al. \cite{DBLP:conf/iui/PuertaE99} noted that mapping abstract models to concrete interface elements remains a key bottleneck. 
Although these methods ensure consistency in interface development, their reliance on static, pre-defined models imposes a high barrier to entry.
%基于规格的 UI 生成进一步从功能、参数或分析规格推导可视控件[26-28]。NL4DV[8]将自然语言查询转化为属性、任务与 Vega-Lite 规格，SpecifyUI[35]则从参考界面中提取结构化表示并支持分层定向编辑。这类工作提升了界面生成的自动化与可控性，但仍高度依赖详尽的前置规格输入。
Specification-based UI generation further derives visual controls from functional, parametric, or analytical specifications \cite{10.1145/3332165.3347944, 10.1145/571985.572008, 10.1145/964442.964507}. NL4DV \cite{9222342} transforms natural language queries into attributes, tasks, and Vega-Lite specifications, while SpecifyUI \cite{DBLP:journals/corr/abs-2509-07334} extracts structured representations from reference interfaces and supports hierarchical targeted editing. While these works enhance the automation and controllability of interface generation, they still rely heavily on a priori specifications.
%随着 LLM 的发展，研究者开始从自然语言生成交互界面和前端代码[29-34]。相关工作可生成任务特定界面或定制网页[29,30]；BISCUIT[31]在代码生成前插入临时 UI，帮助用户通过控件引导生成；近期工作通过语义中间层、设计系统约束或层级澄清控件增强可控性与歧义处理[32-34]。这类工作证明了 LLM 生成界面的潜力，但其输出稳定性和可解释性仍有限，且难以处理特定领域复杂的内部结构与约束关系。
With the development of LLMs, researchers have begun to generate interactive interfaces and front-end code from natural language~\cite{DBLP:journals/corr/abs-2508-19227,leviathan2026generativeuillmseffective,10714542,10.1145/3772318.3791966, 10.1145/3772363.3798616, 10.1145/3746059.3747686, ye2024generative}. 
% Existing work can generate task-specific interfaces or customized webpages \cite{DBLP:journals/corr/abs-2508-19227, leviathan2026generativeuillmseffective}; BISCUIT \cite{10714542} inserts temporary UIs before code generation to help users steer synthesis through controls; and recent work enhances controllability through semantic intermediate layers, design-system constraints, or hierarchical clarification widgets \cite{10.1145/3772318.3791966, 10.1145/3772363.3798616, 10.1145/3746059.3747686}. 
These works demonstrate the potential of LLM-generated interfaces, but their output stability and interpretability remain limited, and they struggle to handle complex internal structures and constraints in specific domains.

%另一类工作关注终端用户在使用过程中塑造界面。Min 等[36]和 SimStep[37]分别支持概览—细节界面定制与教学仿真人机共创。Jelly[38]以可演化任务数据模型驱动 UI 生成，DuetUI[39]通过任务分解、界面描述和双向上下文回路支持用户与 agent 共创任务导向界面。这些工作增强了界面可塑性，但大多面向通用任务或开放式信息探索。
Another line of work focuses on how end users shape interfaces during use. 
Min et al. \cite{10.1145/3706598.3714164} and SimStep \cite{10.1145/3772318.3791514} support overview-detail interface customization and human-AI co-creation of instructional simulations, respectively. 
Jelly \cite{10.1145/3706598.3713285} drives UI generation with an evolvable task data model, and DuetUI \cite{10.1145/3772318.3790441} supports users and agents in co-creating task-oriented interfaces through task decomposition, interface descriptions, and bidirectional context loops. Existing generative UI research has significantly enhanced interface malleability for general-purpose construction and open-ended exploration.
%现有生成式 UI 研究多侧重于通用界面构建或开放域探索，较少关注可视化对象内部的视觉编码、数据变换与组件协同关系。本文聚焦自然语言驱动的图表编辑场景，将生成式界面从通用任务界面的动态生成推进至图表编辑操作空间生成，以支持面向图表语义与组件关系的交互式编辑。
However, these approaches remain insufficiently tailored to the unique complexities of data visualization, particularly the systemic coordination of visual encodings, data transformations, and chart components. 
Our work focuses on natural language-driven chart editing and extends generative interfaces toward interactive chart-editing spaces grounded in chart semantics and component relations.
\section{Overview}

\subsection{Design Requirements}

%为支持真实自然语言图表编辑场景中常见的复合表达、欠指定编辑目标与跨组件关联修改，我们从现有自然语言可视化编辑系统的局限与挑战出发，提炼出以下设计需求。
%Real-world natural-language-driven chart editing often involves compound requests, underspecified goals, and cross-component modifications. Based on these challenges and the limitations of existing natural language-driven chart editing tools, we derive the following design requirements.

{\color{black} We conducted a literature-driven and task-oriented analysis of prior work on chart editing, visualization NLIs, and generative user interfaces. By examining how representative systems handle compound requests, underspecified intents, hierarchical refinement, and cross-component coordination, we summarized recurring limitations and translated them into the following design requirements.}

%DR1.将复合自然语言编辑请求结构化为可执行子意图。
%用户的自然语言请求往往同时包含多个编辑目标，而不是单一操作。例如，“换成更协调的蓝色主题，并突出最高值”同时涉及全局风格调整、目标数据识别、视觉强调和可能的注释添加。系统应能够将这类复合请求拆解为多个子意图，并描述每个子意图对应的编辑目标、作用范围和相关对象，为后续控件生成和图表更新提供依据。
\textbf{DR1. Structure compound natural language editing requests into executable sub-intents.} Users' natural language requests often contain multiple editing goals rather than a single operation. For example, ``change to a more harmonious blue theme and highlight the maximum value'' involves global style adjustment, target data identification, visual emphasis, and potentially annotation generation. The system should decompose such compound requests into multiple sub-intents and describe the editing target, scope, and related objects of each sub-intent, providing the basis for subsequent widget generation and chart updates.

%DR2.将欠指定编辑请求转化为可控的编辑空间。
%用户有时会以高层或模糊方式表达编辑目标，而不会明确指定编辑对象、属性与参数。例如，“换个颜色主题”并未说明需要调整哪些对象、采用何种配色方案。系统应根据用户的高层意图，确定相关的可编辑对象、编辑维度与参数范围，将欠指定请求转化为可探索、可调整的编辑空间。
\textbf{DR2. Transform underspecified editing requests into controllable editing spaces.} Users may express editing goals in high-level or ambiguous ways without explicitly specifying target objects, attributes, or parameters. For example, ``change the color theme'' does not indicate which objects should be modified or which color scheme should be used. The system should determine relevant editable objects, editing dimensions, and parameter ranges based on the user's high-level intent, thereby transforming an underspecified request into an explorable and adjustable editing space.

%DR3.支持从全局到局部的层级式调整。
%一些编辑请求暗含全局意图，例如调整整体布局或修改颜色主题，但用户通常不会明确说明每个图表组件应如何变化。系统应支持从全局编辑方向到局部参数细化的层级式操作，使用户能够先比较和确定整体方案，再进一步检查和调整具体参数。通过这种方式，用户可以在编辑效率和细粒度调整之间取得平衡。
\textbf{DR3. Support hierarchical adjustment from global editing to local refinement.} Some editing requests imply global intents, such as adjusting the overall layout or modifying the color theme, but users usually do not specify how each chart component should change. The system should facilitate hierarchical operations, spanning from high-level editing intents to fine-grained parameter refinement. 
Users should be able to first compare and select an overall scheme, and then inspect and adjust specific parameters. In this way, users can balance editing efficiency with fine-grained adjustment.

%DR4. 识别编辑目标相关的潜在影响字段。
%图表组件并非彼此孤立，一个编辑目标可能涉及与其存在视觉编码关系、结构关系或表达一致性关系的其他字段。例如，mark 与 legend 需要保持颜色编码一致；新增类别可能影响 marks、legends、颜色映射和类别标签；高亮最高值可能同时涉及 marks、labels 和 annotations。系统应识别这些潜在影响字段并提供原因，为后续跨组件联动控制提供依据。
\textbf{DR4. Identify potentially affected fields related to editing targets.} Chart components are not isolated from one another. A single editing target may involve other fields connected through visual encoding relations, structural relations, or expressive consistency relations. For example, marks and legends need to maintain color-encoding consistency; adding a new category may affect marks, legends, color mappings, and category labels; highlighting the maximum value may involve marks, labels, and annotations simultaneously. The system should identify these potentially affected fields and explain their relevance, providing the basis for subsequent cross-component coordinated controls.

%DR5. 支持关系感知的跨组件联动控制。
%图表编辑请求常常隐含多个相关组件的同步变化。系统应利用图表组件之间的关系，以可检查、可调整的方式支持可控的跨组件联动，从而维护视觉编码一致性、结构完整性和表达一致性。
\textbf{DR5. Support relation-aware cross-component coordinated controls.} Chart editing requests often imply synchronous changes across multiple related components. The system should leverage relationships among chart components to support controllable cross-component coordination in an inspectable and adjustable manner, thereby maintaining visual encoding consistency, structural integrity, and expressive consistency.

\begin{figure}[t]% specify a combination of t, b, p, or h for top, bottom, on its own page, or here
  \centering % avoid the use of \begin{center}...\end{center} and use \centering instead (more compact)
  \includegraphics[width=\columnwidth]{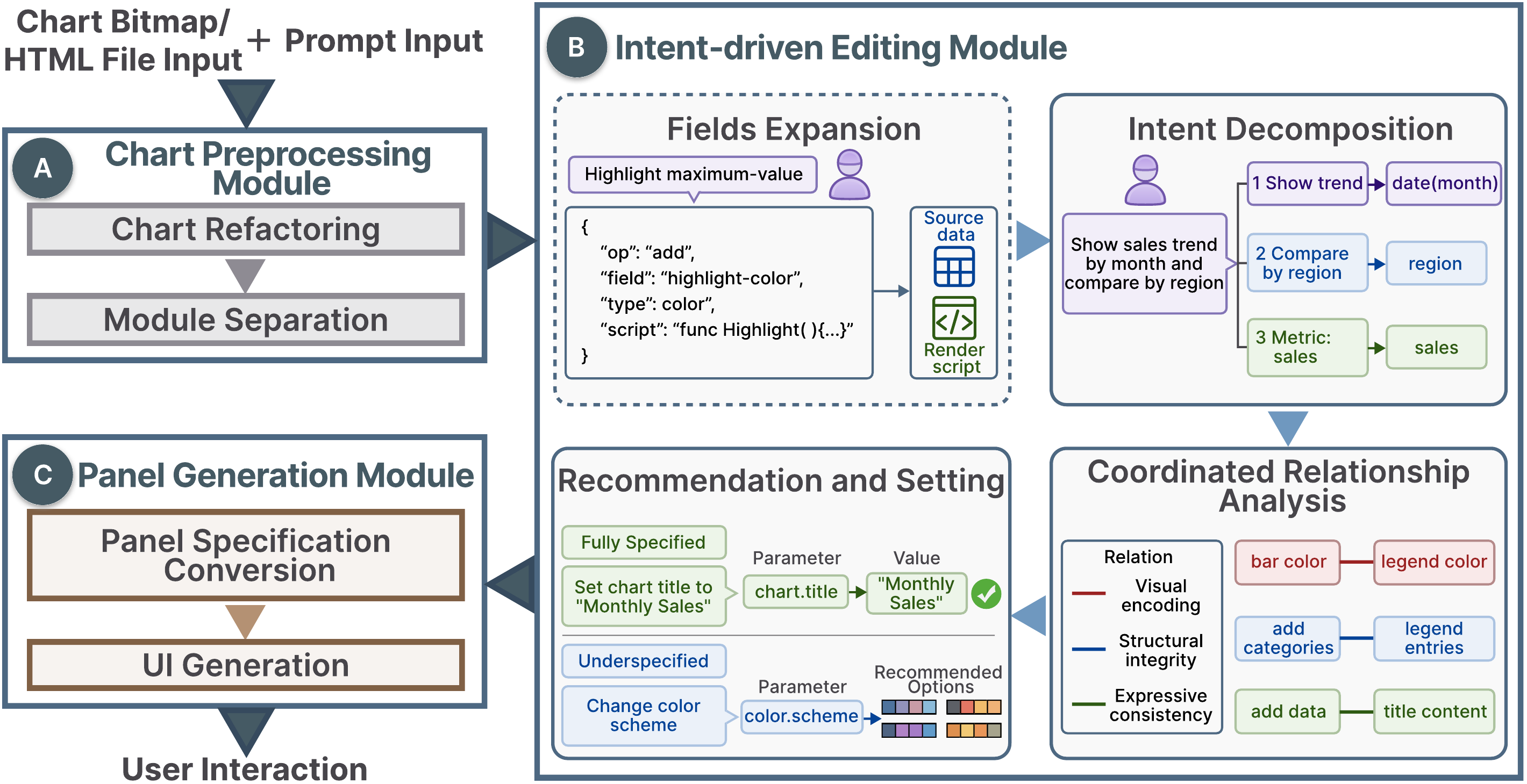}
  \caption{%\ours 的工作流。用户上传图表位图或 HTML 文件并输入自然语言请求。Chart Preprocessing Module (A) 将输入图表重构并拆分为可编辑的数据与渲染逻辑；Intent-driven Editing Module (B) 完成字段扩展、意图分解、关系分析及参数推荐或设定；Panel Generation Module (C) 将结构化编辑信息转换为面板规格并生成交互式 UI，支持用户检查推荐结果、调整参数并实时更新图表。
  	%The workflow of \ours. Users upload a chart bitmap or HTML file and enter a natural language request. The Chart Preprocessing Module (A) performs Chart Refactoring and Module Separation to obtain editable source data and render script; the Intent-driven Editing Module (B) conducts Fields Expansion, Intent Decomposition, Coordinated Relationship Analysis, and Recommendation and Setting; the Panel Generation Module (C) converts the structured editing information through Panel Specification Conversion and UI Generation, enabling users to inspect recommendations, adjust parameters, and update the chart in real time.
    The workflow of \ours. Given a chart (bitmap/HTML) and a NL request, The Chart Preprocessing Module (A) extracts editable data and rendering scripts via Chart Refactoring and Module Separation. The Intent-driven Editing Module (B) then processes the request through Intent Decomposition, Fields Expansion, Coordinated Relationship Analysis, and Recommendation Setting. Finally, the Panel Generation Module (C) translates these structured outputs into interactive UIs.
  }
    \vspace{-3mm}
  \label{fig:workflow}
\end{figure}

\subsection{System Overview}

%为支持自然语言驱动的图表编辑，RAGE-Vis 采用如图 X 所示的三阶段工作流：图表预处理、意图驱动编辑和面板生成。系统以图表图像或 HTML 图表以及用户的自然语言编辑请求作为输入，并生成可交互的图表编辑面板。
For natural language-driven chart editing, we present \ours, which employs a three-stage workflow comprising chart preprocessing, intent-driven editing, and panel generation, as shown in \cref{fig:workflow}. \ours accepts a chart image or HTML source alongside a natural language request, dynamically constructing interactive editing panels.

%首先，图表预处理模块对输入图表进行重构，并将图表中的数据、视觉属性和渲染逻辑组织为可参数化编辑的中间表示，为后续的意图解析、关系分析和面板生成提供基础。
First, the \textbf{Chart Preprocessing Module} (\hyperref[fig:workflow]{\cref*{fig:workflow}(A)}) refactors the input chart and organizes its data, visual attributes, and rendering logic into a parameterized intermediate representation for subsequent intent parsing, relation analysis, and panel generation.

%其次，意图驱动编辑模块将自然语言请求转化为结构化编辑信息。当当前图表表示无法支持目标编辑时，该模块首先扩展图表字段与渲染逻辑。随后，该模块将用户请求拆解为子意图，识别编辑目标及其相关字段，并根据请求的明确程度直接设置参数或生成推荐。
Second, the \textbf{Intent-driven Editing Module} (\hyperref[fig:workflow]{\cref*{fig:workflow}(B)}) transforms the natural language request into structured editing information. It first extends chart fields and rendering logic when the representation cannot support the requested edit. It then decomposes the request into sub-intents, identifies editing targets and related fields, and either sets parameter values or generates recommendations.

%最后，面板生成模块将结构化编辑信息转换为统一的面板规格，并生成对应的交互式控件。通过这些控件，用户可以检查系统推荐结果、调整目标字段和相关字段，并将修改实时应用到图表中。整体而言，该工作流将自然语言编辑请求转化为可检查、可调整的交互式编辑过程，从而在降低代码操作负担的同时支持跨组件一致性维护。
Finally, the \textbf{Panel Generation Module} (\hyperref[fig:workflow]{\cref*{fig:workflow}(C)}) converts the structured editing information into a unified panel specification and generates corresponding interactive controls. Through these controls, users can inspect recommendations, adjust target and related fields, and apply modifications to the chart in real time. Overall, this workflow turns natural language requests into an inspectable and adjustable editing process, reducing the burden of code manipulation while supporting cross-component consistency maintenance.

\section{\ours}

%RAGE-Vis 是一个 Web 应用，专用于基于 D3.js 绘制语言的图表编辑，前端采用 Vue 和 JavaScript 实现，后端采用 Python 实现。
\ours is a web application designed for editing charts authored in D3.js. 
The frontend is implemented with Vue and JavaScript, while the backend is implemented in Python.

\subsection{Cross-component Relations}

\begin{figure*}[t!]
    \centering
    \includegraphics[width=0.85\linewidth]{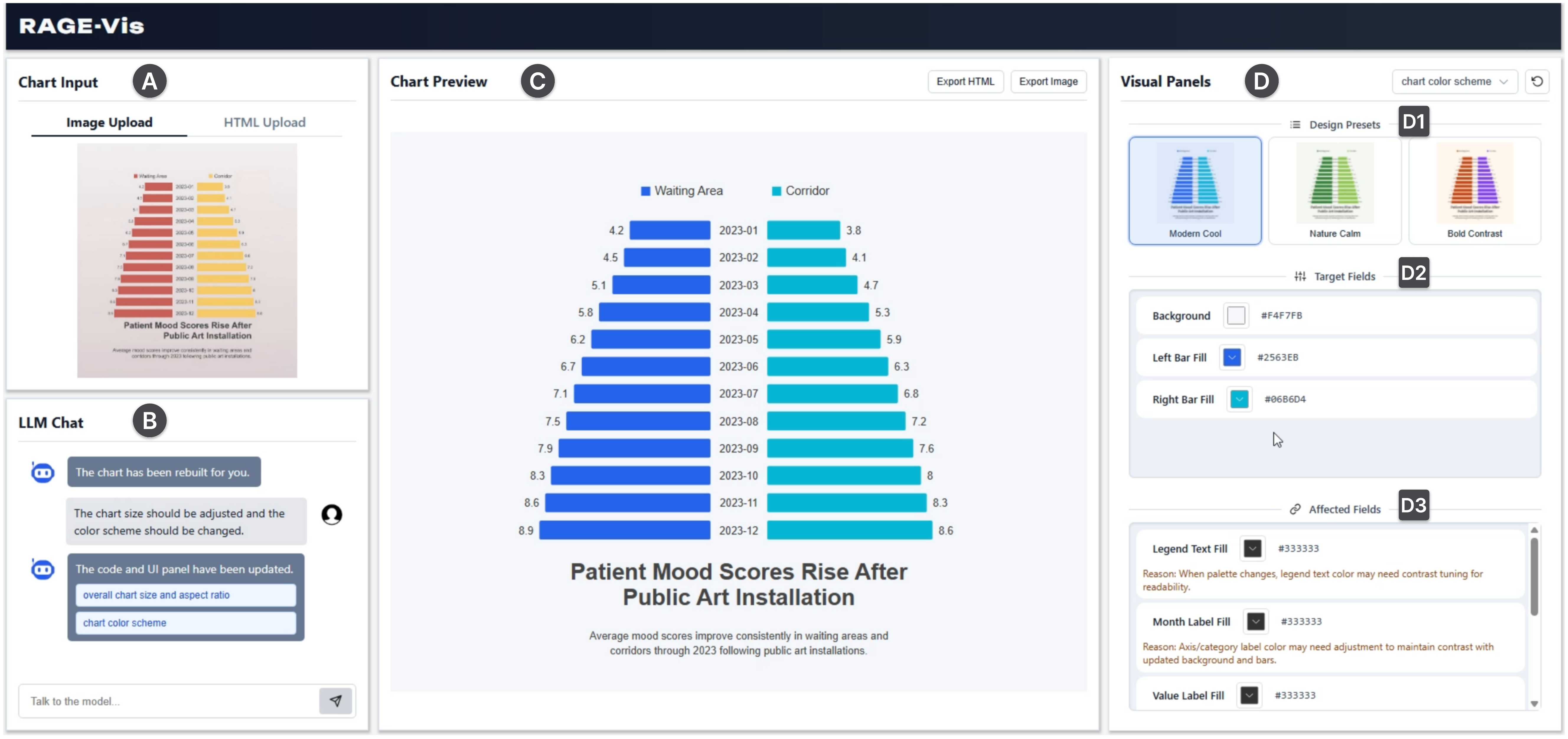}
    \caption{
    %RAGE-Vis 的系统界面，包括 Chart Input (A)、LLM Chat (B)、Chart Preview (C) 和 Visual Panels (D)。
    System interface of \ours, including \textit{Chart Input} (A), \textit{LLM Chat} (B), \textit{Chart Preview} (C), and \textit{Visual Panels} (D).
    }
    \vspace{-3mm}
    \label{fig:system}
\end{figure*}

%为支持图表编辑中的关系感知控制，RAGE-Vis 在参数化中间表示之上对图表内部关联进行建模。具体而言，系统将可编辑字段作为基本分析单元，并将其组织为一种带类型的关系结构。每个字段都关联其在 source data 中的路径、所属图表组件、对应视觉通道、语义角色以及依赖元数据。基于这一字段级表示，系统不仅能够识别用户请求直接命中的字段，还能够进一步发现编辑过程中需要暴露给用户检查的相关字段。在此基础上，我们将图表内部关联划分为三类：视觉编码关系、结构关系和表达一致性关系。
To support relation-aware controls during chart editing, \ours represents internal chart interconnections as typed field-level relations over the intermediate representation. Specifically, the system treats editable fields as the basic analysis unit and organizes them into a typed relation structure. Each field is associated with its source-data path, corresponding chart component, visual channel, semantic role, and dependency metadata. This representation enables the system to identify not only the fields directly targeted by a user request, but also additional related fields that need to be exposed. Based on this field-level representation, we categorize internal chart interconnections into three types: \textbf{visual encoding relations, structural relations, and expressive consistency relations.} 

\subsubsection{Visual Encoding Relations}
%视觉编码关系用于连接那些共同实现同一“数据到视觉映射”的字段。实际实现中，当多个字段共享同一编码来源、数据键或语义映射角色时，系统便建立这一关系。典型例子包括：同时作用于 chart marks 与 legend swatches 的类别颜色。当某个处于此类关系中的字段被编辑时，系统会继续追踪其他依赖相同编码来源的字段，并将其纳入相关字段分析。通过这种方式，系统能够维护不同图表组件之间的一致映射，避免局部修改后出现图例不匹配、颜色不一致或位置表达误导等问题。
Visual encoding relations connect fields that jointly realize the same data-to-visual mapping. In practice, this relation is established when multiple fields share the same encoding source, data key, or semantic mapping role. Typical examples include category colors shared by chart marks and legend swatches. When a field involved in such a relation is edited, the system traces other fields that depend on the same encoding source and includes them in related-field analysis. In this way, the system helps maintain consistent mappings across chart components and avoids mismatched legends, inconsistent colors, or misleading positional cues after local edits.

\subsubsection{Structural Relations}
%结构关系用于连接那些由图表构造逻辑建立起来的字段。在我们的实现中，它主要包括数据生成依赖和布局依赖两类。前者描述哪些视觉元素由哪些数据结构或参数组生成，后者描述某个组件的空间配置如何约束其他组件，例如标题尺寸影响图例位置，图例高度影响可用绘图区范围。基于这些依赖，当编辑涉及数据插入、删除、重排、图例组织或布局调整时，系统能够进一步检索相关字段，从而保证图表在结构组成和空间布局上保持完整与协调。
Structural relations connect fields through chart construction logic. In our implementation, they mainly include data generation dependencies and layout dependencies. Data generation dependencies describe which visual elements are generated from which data structures or parameter groups, while layout dependencies describe how the spatial configuration of one component constrains others, such as title size affecting legend position or legend height affecting the available plotting region. Based on these dependencies, the system can retrieve additional related fields when edits involve data insertion, deletion, reordering, legend arrangement, or layout adjustment, thereby preserving structural completeness and spatial coherence.

\subsubsection{Expressive Consistency Relations}
%表达一致性关系更强地依赖于当前编辑意图。它描述的是多个字段需要协同工作，以共同实现同一表达目标，例如强调、比较、解释或叙事澄清。为了将这类关系落地，系统首先推断用户请求隐含的表达目标，再实例化与该意图对应的关系模板，以检索出应共同支撑该目标的字段。因此，表达一致性并不被建模为完全固定的静态依赖，而是由当前编辑任务动态激活的关系模式。这使得系统能够在分析强调效果和视觉叙事层面维持整体连贯性，而不仅仅停留在外观或结构层面的同步。
Expressive consistency relations are more strongly conditioned on the current editing intent. They capture cases where multiple fields need to work together to fulfill the same communicative goal, such as emphasis, comparison, explanation, or narrative clarification. To operationalize this relation, the system first infers the expressive goal implied by the user request and then instantiates an intent-specific relation template to retrieve the fields that should jointly support that goal. Therefore, expressive consistency is not modeled as a fully fixed dependency, but is represented as a dynamic relation pattern activated by the current editing task. This allows the system to preserve coherence at the level of analytical emphasis and visual narrative, rather than only at the level of appearance or structure.

\subsection{System Implementation}
%RAGE-Vis 整体采用基于多模态模型与规则结合（multimodal-model and rule-based）的实现方式。其中，多模态模型主要负责图表理解、代码生成、意图解析与参数推荐等语义相关任务，规则系统则用于字段映射、关系分析、结构约束以及界面生成等确定性处理。系统核心能力基于 GPT-5.3 Codex 模型实现。为保证不同模块之间的数据一致性与可组合性，系统所有阶段的输入与输出均采用格式化 JSON 数据进行组织。
\ours adopts an implementation strategy that combines multimodal models with rule-based processing. The multimodal model is primarily responsible for semantic tasks, including chart understanding, code generation, intent parsing, and parameter recommendation. The rule-based component handles deterministic processing, such as field mapping, relation analysis, structural constraints, and interface generation. The core capabilities of the system are built on the GPT-5.3-Codex model. To ensure data consistency and composability across modules, all inputs and outputs throughout the pipeline are organized as structured JSON objects.
{\color{black}We provide additional implementation details in supplementary material, including the internal representation, bitmap-to-editable-chart preprocessing, editable field extraction, intent decomposition and relation inference, field expansion, recommendation and panel generation, and validation rules.}

\subsubsection{Chart Preprocessing}
%系统首先将用户输入的图表图片或 HTML 代码统一转换为面向参数编辑的 HTML 表示。对于图表图像输入，系统通过图表重构生成完整的 HTML 图表代码；对于 HTML 图表输入以及通过重构得到的 HTML 图表代码，系统进一步分析原始 D3 渲染逻辑，将原本分散在绘制流程中的数据与视觉属性统一抽取出来，并组织为可编辑的参数结构。 
The system first converts user-provided chart images or HTML code into a unified HTML representation for parameter editing. For chart image inputs, the system reconstructs a complete HTML chart from the image. For HTML chart inputs, as well as reconstructed HTML charts, the system further analyzes the original D3 rendering logic and extracts the data and visual attributes that were previously scattered throughout the rendering process, reorganizing them into an editable parameter structure.

%在此基础上，系统将完整图表进一步拆分为 source data 与 render script 两部分。其中，source data 用于存储图表中可被编辑的内容与参数，包括原始数据、派生字段、视觉编码、样式属性、文本标注以及字段与图表组件之间的映射关系；render script 则负责读取 source data 中的字段并生成对应的 SVG/HTML 元素，同时在参数发生变化时完成图表重渲染。通过这种拆分方式，系统将图表的可编辑内容与渲染逻辑解耦，为后续基于意图的字段定位、关系分析与界面生成提供了统一基础。
Based on this representation, the complete chart is further decomposed into two parts: source data and render script. The source data stores all chart content and parameters that can be edited, including raw data, derived fields, visual encodings, style attributes, text annotations, and mappings between fields and chart components. The render script reads fields from the source data and generates the corresponding SVG/HTML elements, while re-rendering the chart whenever parameters are changed. This decomposition decouples editable chart content from rendering logic and provides a unified basis for subsequent intent-driven field localization, relation analysis, and interface generation.

\subsubsection{Intent-Driven Editing}
%在获得可编辑的图表表示后，系统通过四个步骤将用户的自然语言请求转化为结构化编辑结果：字段扩展、意图分解、关系感知字段扩展，以及推荐或直接设置。
After obtaining an editable chart representation, the system transforms the user’s natural-language request into structured editing results through four steps: Fields Expansion, Intent Decomposition, Coordinated Relationship Analysis, Recommendation and Setting.

%字段扩展。RAGE-Vis 结合用户意图与当前 source data，判断现有图表表示是否已经包含支持目标编辑所需的参数结构。若当前表示不足以支撑该编辑，系统会扩展 source data，并对 render script 进行增量修改，使新增参数能够参与后续渲染。
\textbf{Fields Expansion.} By jointly considering the user intent and the current source data, \ours determines whether the existing chart representation already contains the parameter structure required to support the requested edit. If the current representation is insufficient, the system expands the source data and incrementally modifies the render script, so that the newly introduced parameters can participate in subsequent rendering.

%意图分解。RAGE-Vis 将用户请求分解为多个结构化子意图，并为每个子意图在 source data 中定位相应的 target fields。在我们的实现中，Target Fields 仅包含与当前编辑意图直接对应的属性，从而使生成的控件始终聚焦于用户明确提出的编辑目标。
\textbf{Intent Decomposition.} \ours decomposes the user request into multiple structured sub-intents and identifies the corresponding target fields in the source data for each sub-intent. In our implementation, Target Fields contain only the attributes that directly correspond to the current editing intent, so that the generated controls remain focused on the explicit editing target. {\color{black} During intent decomposition, RAGE-Vis records uncertain aspects of each sub-intent, including possible editing targets, operation scopes, and parameter values. If the target, scope, and value are explicitly specified, the system directly updates the corresponding parameters. If the request is underspecified but the editing dimension can be inferred, the system generates candidate presets or parameter options. If multiple plausible interpretations exist, RAGE-Vis keeps these alternatives inspectable in the generated panel rather than committing to a single interpretation. In this way, ambiguity is handled through a mixed-initiative process in which the system proposes structured alternatives and users resolve uncertainty through selection and refinement.}

%联动关系分析。RAGE-Vis 以这些 target fields 为起点，沿视觉编码关系、结构关系和表达一致性关系执行关系感知的字段扩展，检索其他相关字段，并将这些扩展得到的字段统一组织为 Affected Fields。在我们的实现中，无论依赖关系较强还是较弱，只要能够帮助用户理解更广泛的跨组件影响，并判断是否需要进一步联动调整，都会被纳入 Affected Fields。它们之间的区别不体现在是否展示，而体现在说明方式与优先级上。因此，每个 affected field 都会附带简短原因，用于解释它为何与当前编辑目标相关。
\textbf{Coordinated Relationship Analysis.} {\color{black}Starting from the target fields identified for each sub-intent, \ours searches the field-level relation structure to find additional fields that may need to be inspected or edited together with the target. The search follows three relation types: visual encoding relations, structural relations, and expressive consistency relations.} The retrieved fields are uniformly organized into Affected Fields. In our implementation, {\color{black}the system retrieves relevant fields and determines dependencies based on a relation template file. However,} both stronger and weaker dependencies are included in Affected Fields, since both can help users inspect broader cross-component implications and decide whether further coordinated adjustment is needed. Their difference is reflected not in whether they are shown, but in how they are explained and prioritized. Each affected field is therefore accompanied by a brief reason explaining why it is related to the current editing target.

%推荐和设置。RAGE-Vis 根据用户请求的明确程度来确定参数值。若用户同时明确指出编辑目标和期望取值，系统将直接设置相应参数；对于欠明确或探索式请求，系统则结合当前图表状态与用户意图生成多组推荐参数值，作为后续交互中的候选方案。通过这一过程，自然语言请求被转化为可执行、可解释的结构化编辑表示。
\textbf{Recommendation and Setting.} \ours determines parameter values according to the specificity of the user request. If the user explicitly specifies both the editing target and the desired value, the system directly sets the corresponding parameters. For underspecified or exploratory requests, the system instead generates multiple recommended parameter values based on the current chart state and user intent, providing candidate options for subsequent interaction. Through this process, natural-language requests are transformed into executable and interpretable structured editing representations.

\subsubsection{Panel Generation}

%系统将前一阶段得到的结构化编辑结果转换为统一的面板规格，使其能够被前端界面直接解析与渲染。针对不同类型的编辑字段，系统映射生成相应的交互控件，例如将数值类字段映射为输入框，将颜色类字段映射为颜色选择器，将文本类字段映射为文本输入框，将数据集合类字段映射为表格控件；对于与推荐相关的字段，系统会生成预设选项，并以预览图的形式供用户选择，使用户能够在后续编辑前快速比较不同设计方案。
The system converts the structured editing results produced in the previous stage into a unified panel specification, so that they can be directly parsed and rendered by the front-end interface. Different types of editable fields are mapped to corresponding interaction widgets: numeric fields are mapped to input boxes, color fields to color pickers, text fields to text inputs, and data collection fields to table-based controls. For fields associated with recommendations, the system generates preset options and presents them as preview thumbnails for user selection, enabling users to compare alternative designs at a glance before proceeding with subsequent edits.

%在界面组织上，系统首先按照复合请求分解得到的子意图对目标字段进行分组展示，使每组控件对应一个相对完整的局部编辑任务。与目标字段不同，受影响字段会附带对应的影响理由，用于说明这些字段为何会随着目标编辑发生联动变化。用户在界面中的参数修改将实时回传至 source data，再由 render script 完成图表重绘。通过这一过程，系统实现了从自然语言编辑意图到可操作参数面板的转换，使用户能够以参数化、可视化的方式完成后续编辑。
At the interface level, the system first groups target fields according to the sub-intents obtained from decomposing the compound request, so that each group of controls corresponds to a relatively self-contained local editing task. Unlike target fields, affected fields are accompanied by corresponding dependency reasons, explaining why they change together with the target edit. User modifications made through the interface are fed back to the source data in real time, after which the render script re-renders the chart. Through this process, the system transforms natural-language editing intent into an operable parameter panel, enabling users to perform subsequent edits in a parameterized and visual manner.

\begin{figure*}[t!]
    \centering
    \includegraphics[width=0.85\linewidth]{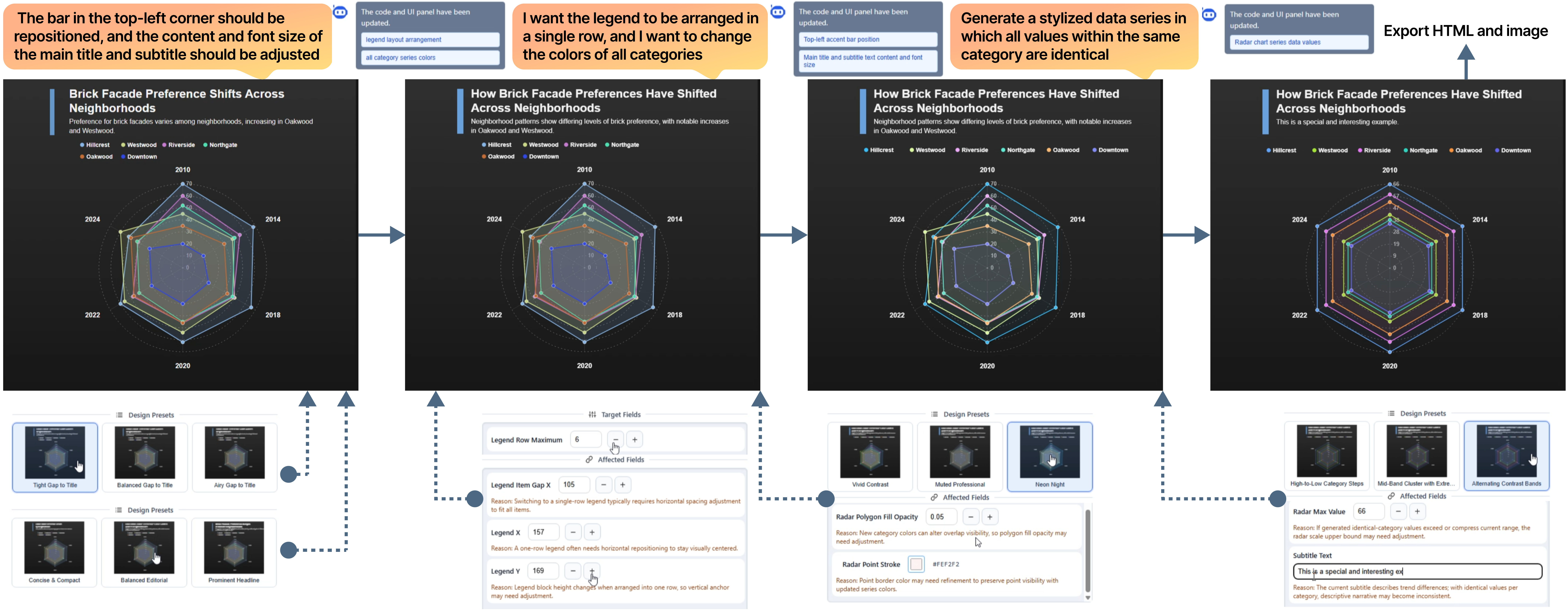}
    \caption{
    %案例二：RAGE-Vis 支持多类别雷达图上的关系感知编辑。 系统为标题/装饰元素调整、图例与颜色编辑以及特色数据生成提供可视化控件，并支持用户协调相关组件。
    \textbf{Case II: \ours supports relation-aware editing of a multi-category radar chart.} The system generates visual controls for title/decorative adjustment, legend and color editing, and stylized data generation while enabling users to coordinate related components.
    }
    \vspace{-3mm}
    \label{fig:case2}
\end{figure*}

\subsection{User Interface}

%图 X 展示了 RAGE-Vis 的系统界面。界面由四个主要面板组成：Chart Input（A）、LLM Chat（B）、Chart Preview（C）和 Visual Panels（D），支持用户完成从图表导入、自然语言请求提交、结果预览到参数调整的完整编辑流程。
\cref{fig:system} shows the interface of \ours. The interface consists of four main views: \textit{Chart Input}, \textit{LLM Chat}, \textit{Chart Preview}, and \textit{Visual Panels}, supporting the complete editing workflow from chart import and natural language request submission to result preview and parameter adjustment.

%Chart Input（A）支持用户上传图表图像或导入已有 HTML 图表。LLM Chat（B）用于接收用户的自然语言编辑请求，用户既可以输入目标明确的修改指令，也可以以高层或欠指定方式表达编辑目标，而无需预先指定所有相关对象、属性和参数（DR2）。当请求涉及多个编辑目标时，系统会生成一个或多个对应的 Visual Panels，使用户能够分别检查和调整不同编辑任务（DR1）。Chart Preview（C）同步展示当前图表状态，帮助用户检查编辑结果，并判断是否需要通过 Visual Panels 进一步进行局部细化（DR3）。导出按钮支持用户将编辑后的图表保存为图片或 HTML。
\textit{Chart Input} (\hyperref[fig:system]{\cref*{fig:system}(A)}) supports uploading chart images or importing existing HTML charts. \textit{LLM Chat} (\hyperref[fig:system]{\cref*{fig:system}(B)}) receives users' natural language editing requests, including clearly specified editing instructions as well as high-level and underspecified goals, without requiring users to predefine all related objects, attributes, and parameters (\textbf{DR2}). When a request involves multiple editing targets, the system generates one or more corresponding Visual Panels, enabling users to inspect and adjust different editing tasks separately (\textbf{DR1}). \textit{Chart Preview} (\hyperref[fig:system]{\cref*{fig:system}(C)}) synchronously presents the current chart state, helping users inspect editing results and determine whether further local refinement is needed through the Visual Panels (\textbf{DR3}). Export buttons allow users to save the edited chart as an image or HTML.

%Visual Panels（D）是用户检查和调整编辑结果的主要区域。每次自然语言请求都会生成对应的编辑面板，并可通过顶部下拉列表进行切换，便于用户返回先前请求继续调整。面板内部包含 Design Presets、Target Fields 和 Affected Fields 三个部分。对于欠指定或探索性请求，Design Presets（D1）提供多个候选方案，使用户能够比较并选择整体编辑方向；Target Fields（D2）则呈现与当前编辑目标直接相关的字段和控件，支持用户进一步调整具体参数。通过这种层级组织方式，RAGE-Vis 将自然语言编辑请求转化为可检查、可操作的编辑空间，支持用户从整体调整逐步进入局部细化（DR2, DR3）。
\textit{Visual Panels} (\hyperref[fig:system]{\cref*{fig:system}(D)}) are the main area for inspecting and adjusting editing results. Each natural language request generates a corresponding editing panel, which can be revisited through the top drop-down list for further adjustment. Each panel contains three parts: Design Presets, Target Fields, and Affected Fields. For underspecified or exploratory requests, Design Presets (\hyperref[fig:system]{\cref*{fig:system}(D1)}) provide multiple candidate schemes, allowing users to compare and select an overall editing direction. Target Fields (\hyperref[fig:system]{\cref*{fig:system}(D2)}) present fields and controls directly related to the current editing target, supporting further adjustment of specific parameters. Through this hierarchical organization, \ours transforms natural language editing requests into an inspectable and operable editing space, supporting users in moving from overall adjustment to local refinement (\textbf{DR2}, \textbf{DR3}). {\color{black}In this sense, the Visual Panels also serve as a disambiguation interface. Design Presets expose alternative high-level interpretations, Target Fields reveal the parameters that the system assumes to be directly relevant, and Affected Fields show possible cross-component implications. Users can therefore verify and refine the system interpretation before committing to further edits.}

%对于涉及跨组件关联的请求，Affected Fields（D3）补充呈现除直接编辑目标外可能受到影响的相关字段，并为每个字段提供简要原因（DR4）。这些说明将潜在的间接影响显式呈现给用户，帮助用户判断是否需要进一步调整。通过将相关字段组织为可检查、可调整的控件，RAGE-Vis 使用户能够以可控方式完成跨组件联动更新，从而维护视觉编码一致性、结构完整性和表达一致性（DR5）。
For requests involving cross-component relationships, Affected Fields (\hyperref[fig:system]{\cref*{fig:system}(D3)}) additionally present related fields that may be affected beyond the direct editing targets, together with brief reasons for each field (\textbf{DR4}). These explanations make potential indirect effects explicit and help users decide whether further adjustments are needed. By organizing related fields as inspectable and adjustable widgets, \ours enables users to coordinate cross-component changes while maintaining visual encoding consistency, structural integrity, and expressive consistency (\textbf{DR5}).

\section{Evaluation}

% 为评估 RAGE-Vis 的有效性，我们结合案例研究和用户研究，考察其在欠指定请求处理、候选方案探索、相关字段发现与跨组件一致性维护方面的支持能力。
To evaluate the effectiveness of \ours, we combine case studies and a user study to examine its support for underspecified request handling, alternative exploration, related-field discovery, and cross-component consistency maintenance.

\subsection{Case Study}
{\color{black} We selected two case-study charts to cover common inputs and different structural complexity levels. Case Study I uses a bitmap chart image to examine reconstruction and parameterized editing, while Case Study II uses an HTML chart to examine relation-aware editing with richer structural and encoding dependencies. These cases illustrate representative scenarios involving underspecified requests and cross-component dependencies.}

\subsubsection{Case Study I}
%为了展示 RAGE-Vis 如何支持关系感知的图表编辑，我们选取了一个关于患者情绪评分变化的镜像柱状图作为案例。用户首先上传原始图表图像，系统对其进行重建，生成对应的 HTML 代码，并渲染出可编辑的图表结果。在此基础上，用户提出三类编辑需求：尺寸与颜色风格调整、数据扩展并高亮最大值，以及数据顺序反转。
To demonstrate how \ours supports relation-aware chart editing, we use a mirrored bar chart showing changes in patient mood scores as a case study. The user first uploads the original chart image. The system reconstructs it into corresponding HTML code and renders it as an editable chart instance. Based on this reconstructed chart, the user issues three types of editing requests: adjusting the chart size and color style, extending the data while highlighting the maximum value, and reversing the data order (\cref{fig:teaser}).

%首先，用户提出“图表整体大小需要调整，并且需要换一种颜色风格”。系统将其拆解为“整体大小调整”和“颜色风格替换”两个子意图，给出两个对应意图的控件组并分别生成多套可选方案。当用户调整图表大小时，图例位置、中心月份标签位置以及标题和副标题的排布也会联动变化，避免图表放缩后出现拥挤或失衡，这体现了 结构关系。切换配色方案会直接改变背景颜色和两侧的条形颜色，同时，基于视觉编码一致性，系统还会为图例和文本元素提供颜色控制。
First, the user requests that ``\textit{The chart size should be adjusted, and the color style should be changed.}'' The system decomposes this request into two sub-intents, overall size adjustment and color style replacement, and generates two corresponding control groups, each with multiple candidate options. When the user adjusts the chart size, the legend position, the central month-label positions, and the layout of the title and subtitle are updated accordingly, preventing crowding or imbalance after resizing; this reflects the handling of structural relations. Switching the color scheme directly changes the background color and the bar colors on both sides, while the system, based on visual encoding consistency, additionally provides color controls for the legend and text elements.

\begin{figure*}[t]
    \centering
    \includegraphics[width=0.9\linewidth]{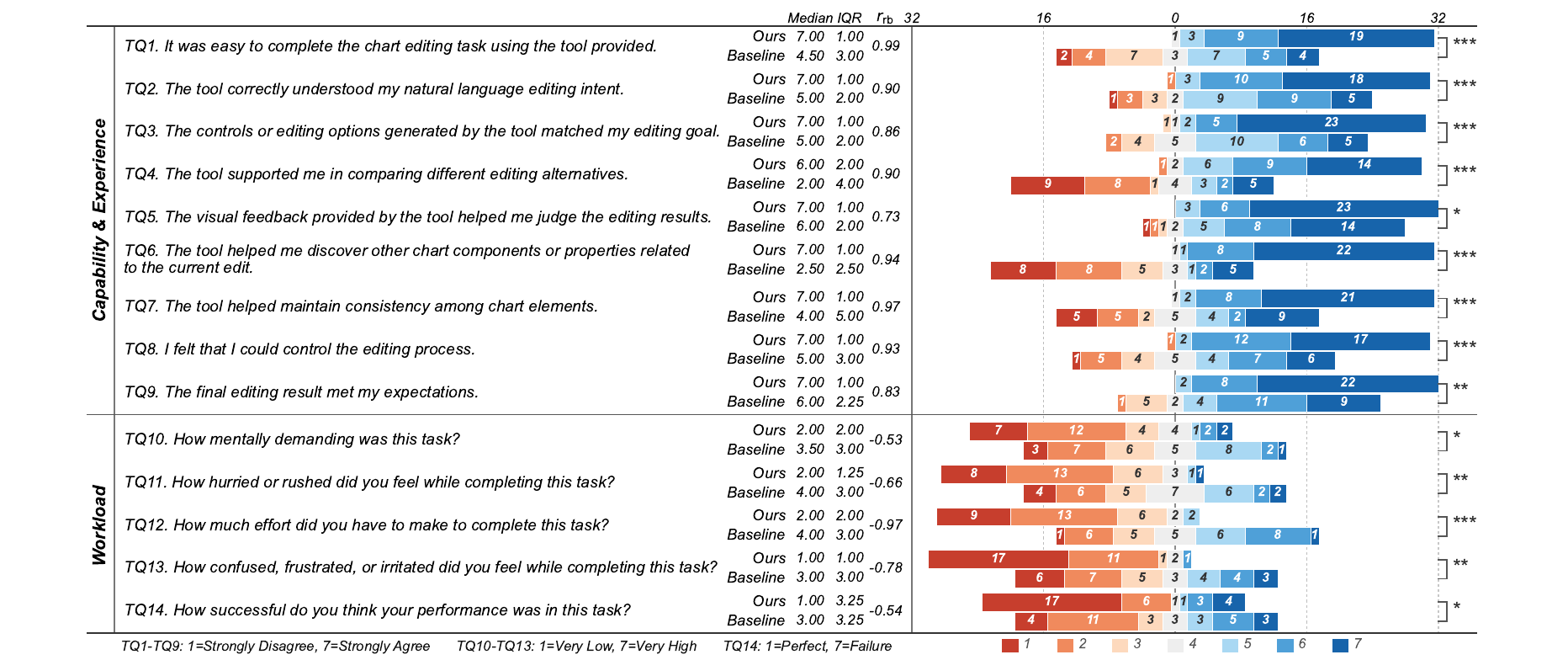}
    %\caption{
    %Post-task questionnaire results, together with the average and standard deviation values and the rating distributions.
    %TQ1--TQ9 measured task experience and perceived system capability, while TQ10--TQ14 measured subjective workload using modified NASA-TLX measures.
    %Significance markers indicate differences between \ours and the baseline (*: $p<.05$, **: $p<.01$, ***: $p<.001$).
    %}
    \caption{
    Post-task questionnaire results, together with the \textcolor{black}{median, interquartile range (IQR), signed rank-biserial effect size ($r_{rb}$),} and rating distributions.
    TQ1--TQ9 measured task experience and perceived system capability, while TQ10--TQ14 measured subjective workload using modified NASA-TLX measures.
    \textcolor{black}{Significance markers indicate differences between \ours and the baseline based on two-sided Wilcoxon signed-rank tests, with $p$-values adjusted using the Holm method (*: adjusted $p<.05$, **: adjusted $p<.01$, ***: adjusted $p<.001$).}
    }
    \label{fig:task-questionnaire}
    \vspace{-3mm}
\end{figure*}

%随后，用户提出“我想要添加 2024-01 的数据，并且高亮最高值”。系统将其拆解为“添加数据”和“新增高亮”两个子意图。系统为图表添加数据的同时，由于原图的副标题说明了时间范围，所以系统进一步给出了副标题文本修改控件，使其与图中数据范围保持一致，这对应 表达一致性关系 的处理。图表在高亮最高值后，强调效果还需要与对应柱形及其标签协调配合，所以系统还给出bar颜色和值文本颜色控件，从而使强调效果更加明显。
Next, the user requests, ``\textit{I want to add the data for 2024-01 and highlight the maximum value.}'' The system decomposes this request into two sub-intents, data addition and new emphasis. While adding the new data, the system also recognizes that the original subtitle specifies the temporal range of the chart, and therefore provides a subtitle text editing control so that the textual description remains consistent with the updated data range; this corresponds to the handling of expressive consistency relations. After highlighting the maximum value, the emphasis effect also needs to remain coordinated with the corresponding bar and its label, so the system further provides controls for the bar color and value-text color, making the highlight more salient.

%最后，用户提出“所有数据倒序排列”。这一需求较为明确，因此系统直接给出倒序后的数据表格。操作完成后，左右两侧柱形顺序、中间月份标签顺序以及对应的位置映射关系同步更新，使倒序后的图表仍保持稳定的镜像比较结构。
Finally, the user requests that ``\textit{I want all data be arranged in reverse order.}'' Because this request is explicit, the system directly provides the reordered data table. Once the operation is applied, the order of the bars on both sides, the order of the central month labels, and the corresponding position mappings are updated together, allowing the reversed chart to preserve a stable mirrored comparison structure.

%总体来看，该案例表明，RAGE-Vis 能够围绕原图中的镜像柱形、中心月份标签、顶部图例、数值标签以及标题副标题，生成结构化、可操作的编辑控件，并能够维护图表属性之间的 视觉编码关系、结构关系 和 表达一致性关系。系统不仅能够针对欠指定或探索性的请求生成推荐方案，还能够针对目标明确的请求提供直接、精确的参数编辑支持。
%Overall, this case shows that \ours can generate structured, operable editing controls around the mirrored bars, central month labels, top legend, value labels, title, and subtitle in the original chart, while preserving the visual encoding relations, structural relations, and expressive consistency relations among chart attributes. The system can not only generate recommendation options for underspecified or exploratory requests, but also provide direct and precise parameter-editing support for well-defined requests.
Overall, this case shows that \ours can coordinate edits across data, bars, labels, legends, and textual descriptions, supporting both recommendation options for underspecified requests and direct parameter editing for explicit requests.

\subsubsection{Case Study II}
%为了进一步展示 RAGE-Vis 在更复杂图表结构上的通用性，我们选取一个多类别雷达图作为第二个案例。系统从用户上传的 HTML 代码出发，重建出可编辑图表，并支持三个的编辑请求：调整装饰条位置并修改主副标题，将图例排成单行并更改所有类别颜色，以及生成一组特色数据。
To further demonstrate the generality of \ours on charts with more complex structures, we use a multi-category radar chart as a second case study (\cref{fig:case2}). Starting from the uploaded HTML chart, the system converts it into an editable chart and supports three requests: repositioning the decorative bar while adjusting the title and subtitle, arranging the legend into a single row while changing category colors, and generating a stylized data series.

%首先，当用户提出调整左上角装饰条位置，并修改主副标题内容与字号时，系统为装饰元素和标题区域生成了与位置、文本和尺寸相关的编辑控件。由于该请求未给出明确数值，系统还提供了推荐方案，供用户后续选择与细化。
First, when the user asks to reposition the top-left decorative bar and modify the title and subtitle content and font size, the system generates editable controls for decorative and title-related elements, including position, text, and size parameters. Because the request does not specify exact values, the system also provides recommendations for subsequent selection and refinement.

%随后，当用户提出将图例排成单行，并更改所有类别颜色时，系统将图例行数相关设置作为直接目标字段，同时补充了与图例布局相关的联动控件。对于类别换色，系统生成多组候选配色方案，并将选定方案一致地应用到图例色块、雷达折线、节点和填充区域，从而保持图例与图表主体之间共享的类别编码一致。
Next, when the user asks to arrange the legend in a single row and change the colors of all categories, the system exposes the legend-row setting as a direct target field and additionally provides related controls for legend layout adjustment. For category recoloring, it generates multiple candidate color schemes and applies the selected scheme consistently to legend swatches, radar polylines, nodes, and filled areas, preserving the shared category encoding between the legend and the chart body.

%最后，当用户要求生成一组同一类别内各维度取值相同的特色数据时，系统将其视为数据生成任务，并基于现有类别结构生成可编辑的雷达图序列数值。为了使生成结果保持在合理显示范围内，系统同时暴露了雷达图最大值这一相关控件；由于批量数据更新还可能改变图表表达含义，系统进一步提供了副标题文本控件。
Finally, when the user requests a stylized data series in which all values within the same category are identical, the system treats this as a data-generation task and produces editable radar-series values based on the existing category structure. To keep the generated values within a reasonable display range, it also exposes the radar maximum value as a related control; because the batch data update may change the chart meaning, the system additionally provides a subtitle text control. Overall, this case shows that \ours remains effective for charts with tighter structural coupling.

%总体来看，该案例表明 RAGE-Vis 能够围绕雷达图结构、图例布局、类别编码、装饰元素和文本描述开展协同编辑。相较于案例一，这个例子进一步说明，面对结构耦合更强、视觉编码共享更明显的图表，系统仍然能够有效支持编辑。
%Overall, this case shows that \ours can support coordinated editing on radar-chart structure, legend layout, category encoding, decorative elements, and textual descriptions. Compared with Case Study I, this example further demonstrates that the system remains effective on charts with tighter structural coupling and more globally shared visual encodings.

\subsection{User Study}

% 为评估 RAGE-Vis 在自然语言图表编辑任务中的有效性与可用性，我们开展了一项被试内用户研究。该研究围绕三个问题展开：RQ1：RAGE-Vis 是否能将模糊或欠指定目标转化为可比较的候选方案；RQ2：Affected Fields 是否能够帮助用户发现与当前编辑目标相关的图表组件或属性，并维护跨组件一致性；RQ3：RAGE-Vis 是否能够减少用户为补充、修正或重新表达编辑意图而额外发起的自然语言请求？
To evaluate the effectiveness and usability of \ours in natural language-based chart editing, we conducted a within-subjects user study. The study focuses on three questions: (1) whether \ours can transform vague or underspecified goals into comparable candidate alternatives; (2) whether Affected Fields can help users discover chart components or properties related to the current editing target and maintain cross-component consistency; and (3) whether \ours can reduce additional natural language requests issued to supplement, correct, or rephrase editing intents.

\subsubsection{Study Design}

% 我们共招募 16 名具有不同程度的数据可视化和图表编辑经验的参与者，并采用被试内设计。研究任务包含两个图表编辑场景，分别基于折线图和堆叠面积图，覆盖样式调整、数据修改、颜色映射、文字编辑等需求。参与者需要参考目标图，将当前图表编辑得尽可能接近目标结果。
We recruited {\color{black}32} participants with varying levels of experience in data visualization and chart editing, and adopted a within-subjects design. {\color{black}The participants, comprising students and researchers, were aged 19 to 30 years (M = 24.75, SD = 2.89). They reported moderate familiarity with data visualization (M = 3.19, SD = 1.38) and chart editing (M = 3.12, SD = 0.81), whereas their familiarity with D3.js was relatively lower and exhibited greater variance (M = 2.38, SD = 1.41).} 
The study included {\color{black}four} chart editing scenarios based on {\color{black}a line chart, a stacked area chart, a scatter plot chart and a pie chart}, covering style adjustment, data modification, color mapping, text editing, and other editing requirements. \textcolor{black}{Participants were asked to edit the provided original chart to match the target chart as closely as possible. Detailed experimental procedures and task designs are provided in the supplementary material.}

% 为降低重复任务带来的学习效应，每位参与者在完整系统和受控基线两个条件下分别完成不同的图表编辑任务。完整系统包含 Design Presets 和 Affected Fields；受控基线移除这两个部分，仅生成与当前编辑目标相关的参数面板，作为 RAGE-Vis 的消融版本考察二者的作用。两个任务与两个系统条件形成四个任务-条件组合，每个组合包含 8 个数据点，共 32 个数据点。
%To reduce the learning effect caused by repeating the same task, each participant completed different chart editing tasks under the full system and control condition. 
% 为减少任务特定的学习效应，每位参与者只编辑每张图一次，并在不同的图表编辑任务中使用两个系统条件。为减轻顺序效应，系统条件的使用顺序在参与者之间进行了平衡：一半参与者先使用 RAGE-Vis，另一半参与者先使用控制条件。
\textcolor{black}{To reduce task-specific learning effects, each participant completed two different chart editing tasks, using \ours in one task and the control condition in the other. The chart-task assignments were balanced across participants, while all participants used \ours before the control condition.}
The full system included Design Presets and Affected Fields, whereas the control condition removed these two parts and only generated parameter panels related to the current editing target, {\color{black}inspired by the dynamic-widget paradigm of DynaVis \cite{10.1145/3613904.3642639},} serving as an ablated version of \ours to examine their roles. \textcolor{black}{The four tasks and two system conditions formed eight task-condition combinations, with eight data points for each combination and 64 data points in total.}

% 在测量与分析方面，我们记录了任务完成时间、模型累计响应时间、用户主动操作时间，以及用户向模型发出的自然语言请求次数。每个任务结束后，参与者填写任务后问卷；完成全部任务后，参与者填写研究后对比问卷。两类问卷分别用于衡量参与者在每个任务后的体验和两个系统条件之间的主观比较。
For measurement and analysis, we recorded task completion time, accumulated model response time, user active operation time, and the number of natural language requests issued to the model. 
%After each task, participants completed a post-task questionnaire; after completing all tasks, they filled out a post-study comparative questionnaire. The two questionnaires measured participants' experience after each task and their subjective comparison between the two system conditions.
Participants completed two post-task questionnaires to evaluate their immediate experiences, followed by a final post-study questionnaire to subjectively compare the two system conditions.
\textcolor{black}{The post-task questionnaire independently measured each system condition, including workload-related items adapted from NASA-TLX. We analyzed its Likert-scale responses using two-sided Wilcoxon signed-rank tests with Holm-adjusted $p$-values and reported median, IQR, and signed rank-biserial effect sizes ($r_{rb}$).}
\textcolor{black}{Adapting ChartEdit's target-matching evaluation protocol~\cite{zhao-etal-2025-chartedit}, we further evaluated the objective quality of the final edited charts by comparing each participant's final chart with the original chart, the target chart, and the task-specific edit checklist. Each result was scored on target edit completion and preservation of non-target elements, both on a 0--5 scale.}

\subsubsection{Results}
% 日志显示，完整系统能够减少用户在完成任务过程中发起自然语言请求的次数。与受控系统相比，参与者在完整系统下发出的自然语言请求次数显著更少（3.44 vs. 6.19，配对 t 检验，p<0.01），模型累计响应时间也显著降低（2分47秒 vs. 4分55秒，p<0.001）。用户主动操作时间接近（9分11秒 vs. 9分27秒），但完整系统的平均总任务时间更短（11分58秒 vs. 14分23秒）。这说明 Design Presets 和 Affected Fields 将候选方案探索与相关字段检查整合到当前面板中，使用户可以直接完成更多检查和调整，减少了通过额外自然语言请求生成新面板或补充遗漏修改的情况。
The logs show that the full system reduced the number of natural language requests issued during task completion. Compared with the control condition, participants issued significantly fewer natural language requests with the full system \textcolor{black}{(4.28 $\pm$ 1.82 vs. 6.25 $\pm$ 2.50, paired $t$-test, $p<.001$)}, and the accumulated model response time was also significantly lower \textcolor{black}{(4 min 38 s $\pm$ 2 min 46 s vs. 6 min 22 s $\pm$ 2 min 58 s, $p<.01$)}. User active operation time was comparable between the two conditions \textcolor{black}{(9 min 12 s $\pm$ 3 min 32 s vs. 9 min 49 s $\pm$ 3 min 50 s, $p=.444$)}, while the average total task time was shorter with the full system 
(\textcolor{black}{mean $\pm$ SD: 13 min 51 s $\pm$ 5 min 03 s vs. 16 min 11 s $\pm$ 6 min 06 s, $p=.053$}). 

% \begin{table}[t]
% \centering
% \caption{Objective quality scores of final edited charts. EC: target edit completion; NP: preservation of non-target elements.}
% \label{tab:objective-quality}
% \footnotesize
% \setlength{\tabcolsep}{3pt}
% \begin{tabular}{lcccccc}
% \toprule
% & \multicolumn{3}{c}{\ours} & \multicolumn{3}{c}{Baseline} \\
% Chart task & EC & NP & Total & EC & NP & Total \\
% \midrule
% Line & 3.83 & 4.54 & 8.37 & 3.67 & 4.42 & 8.08 \\
% Stacked area & 3.83 & 3.92 & 7.75 & 4.00 & 4.08 & 8.08 \\
% Bubble scatter & 4.00 & 4.46 & 8.46 & 4.75 & 4.88 & 9.63 \\
% Pie & 3.87 & 4.33 & 8.21 & 3.46 & 4.21 & 7.67 \\
% \midrule
% Overall & 3.88 & 4.31 & 8.20 & 3.97 & 4.40 & 8.36 \\
% \bottomrule
% \end{tabular}
% \end{table}

\textcolor{black}{The objective quality evaluation further shows that the reduced interaction cost did not come at the expense of final-chart quality. Across all tasks, \ours achieved a comparable overall quality score to the baseline condition (8.20 vs. 8.36). The two sub-scores also remained close overall, including target edit completion (3.88 vs. 3.97) and preservation of non-target elements (4.31 vs. 4.40). The slightly lower score of \ours may be partly related to the fixed condition order in our study: participants used \ours before the baseline condition, which may have introduced a learning effect that benefited the later baseline tasks. The full scoring results are provided in the supplementary material.}

% 任务后问卷结果进一步支持了上述观察（Fig.~X）。在系统能力与任务体验方面，完整系统在所有 TQ1–TQ9 指标上均高于受控系统，其中支持比较不同编辑方案（TQ4: 6.44 vs. 3.19）、发现相关组件或属性（TQ6: 6.75 vs. 3.31）和维护图表元素一致性（TQ7: 6.75 vs. 4.31）上的差异尤为明显（均 p<0.001）。这些结果表明，完整系统支持用户比较候选方案，并检查当前修改可能影响的相关字段。P4 提到，"候选方案让我可以先看几个可能的方向，再决定具体怎么改，不用一开始就想清楚所有参数。" 这说明 Design Presets 对欠指定请求的价值并不只是自动生成一个结果，而是将模糊目标展开为可比较的编辑空间。与此同时，Affected Fields 也增强了用户对跨组件影响的感知。P11 表示，"它会提醒我图例、标签这些地方也可能要一起看一下，不然我可能只会改图里面最明显的部分。" 在主观负荷方面，完整系统整体获得了更低评分，表明候选方案和相关字段提示并未增加额外负担，反而降低了用户的努力程度和挫败感。
The post-task questionnaire results further support these observations (\cref{fig:task-questionnaire}). In terms of perceived system capability and task experience, the full system received higher ratings than the control condition across all TQ1--TQ9 items. The differences were especially evident in support for comparing editing alternatives \textcolor{black}{(TQ4: median 6.0, IQR 2.00 vs. median 2.0, IQR 4.00, $r_{rb}=.90$)}, discovery of related chart components or properties \textcolor{black}{(TQ6: median 7.0, IQR 1.00 vs. median 2.5, IQR 2.50, $r_{rb}=.94$)}, and consistency maintenance among chart elements \textcolor{black}{(TQ7: median 7.0, IQR 1.00 vs. median 4.0, IQR 5.00, $r_{rb}=.97$)}, all with \textcolor{black}{Holm-adjusted $p<.001$}. 
%These results indicate that the full system supported users in comparing candidate alternatives and inspecting fields that might be affected by the current edit. P4 noted, ``\emph{The candidate options let me first see several possible directions and then decide how to edit, instead of having to know all the parameters from the beginning.}'' This suggests that the value of Design Presets for underspecified requests lies not merely in generating one result, but in expanding vague goals into a comparable editing space. Meanwhile, Affected Fields also improved users' awareness of cross-component effects. P11 stated, ``\emph{It reminded me that things like legends and labels might also need to be checked; otherwise I might only change the most obvious part of the chart.}'' In terms of subjective workload, the full system received lower overall scores, suggesting that candidate options and related-field prompts did not introduce extra burden, but instead reduced users' effort and frustration.
Results indicate the full system effectively supported alternative comparison and affected-field inspection. For underspecified requests, Design Presets expanded vague goals into a comparable design space; as P4 noted, the candidate options let he ``\emph{see several possible directions... instead of having to know all parameters from the beginning.}'' Meanwhile, Affected Fields enhanced awareness of cross-component dependencies, reminding P11 to check peripheral elements like ``\emph{legends and labels}'' rather than just the most obvious parts. Consequently, the full system yielded lower subjective workload scores, demonstrating that these guidance features reduced user effort and frustration without introducing extra cognitive burden.

\begin{figure}[t]
    
    \centering
    \includegraphics[width=\linewidth]{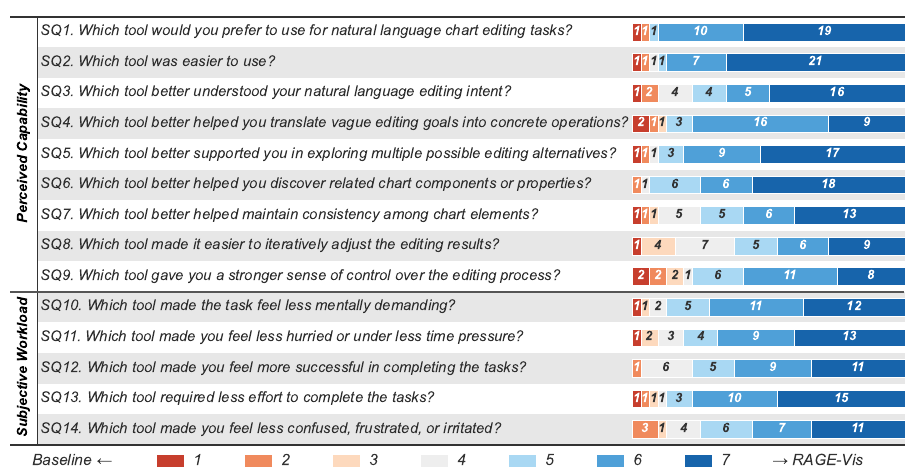}
    \caption{
    Post-study comparative questionnaire results.
    SQ1--SQ9 measured overall preference and perceived system capability, while SQ10--SQ14 measured subjective workload-related comparisons.
    \textcolor{black}{A score of 1 indicates the baseline, 4 indicates a neutral comparative response without preference for either system, and 7 indicates \ours.
    Numbers inside bars indicate participant counts.}
    }
    \label{fig:post-study-questionnaire}
    \vspace{-4mm}
\end{figure}

% 研究后对比问卷显示，参与者整体更偏好完整系统（Fig.~X）。在总体偏好和易用性方面，绝大多数参与者选择了 RAGE-Vis；同时，参与者尤其认可完整系统在探索多个编辑方案（SQ5）、发现相关图表组件或属性（SQ6）、维护图表元素一致性（SQ7）以及降低任务努力程度（SQ13）方面的作用。与受控系统只围绕当前修改目标生成参数面板不同，完整系统通过 Design Presets 和 Affected Fields 将候选方案、目标字段和潜在相关字段组织在同一编辑过程中，使用户能够在一次自然语言请求之后继续检查、比较和调整。P8 提到，"完整系统更像是把后面可能还要问的东西先列出来了，我可以直接在面板里看。" 这也解释了自然语言请求次数和模型等待时间的降低。
%The post-study comparative questionnaire shows that participants generally preferred the full system (\cref{fig:post-study-questionnaire}). In terms of overall preference and ease of use, most participants favored \ours. They particularly recognized the benefits of the full system in exploring multiple editing alternatives (SQ5), discovering related chart components or properties (SQ6), maintaining consistency among chart elements (SQ7), and reducing the effort required to complete tasks (SQ13). Unlike the control condition, which only generated parameter panels around the current editing target, the full system organized candidate alternatives, target fields, and potentially affected fields into the same editing process, allowing users to continue inspecting, comparing, and adjusting after a single natural language request. P8 commented, ``\emph{The full system felt like it listed the things I might need to ask about later, so I could just check them in the panel.}'' This helps explain the reduction in both natural language requests and accumulated model response time.
The post-study questionnaire results reveal a strong overall preference for \ours, particularly regarding ease of use (\cref{fig:post-study-questionnaire}). Participants highly rated the full system for exploring alternatives (SQ5), discovering related components (SQ6), maintaining consistency (SQ7), and reducing effort (SQ13). Unlike the control condition's localized panels, \ours integrates candidate options, target fields, and affected fields into a unified workflow, enabling continuous refinement from a single natural language request. P8 commented, ``\emph{The full system felt like it listed the things I might need to ask about later, so I could just check them in the panel.}'' This effectively explains the observed decrease in both NL requests and accumulated model latency.

\vspace{-2mm}

\subsubsection{Core Pipeline Evaluation}

{\color{black}To further examine the reliability of the core pipeline, we conducted an offline evaluation on the natural-language requests used in the user study. For intent decomposition, two experts manually annotated the expected sub-intents for each request and compared them with the system outputs. \ours achieved an intent-decomposition precision of 82.02\%. For affected-field expansion, the experts judged whether the generated affected fields were necessary for each target edit. The false-positive ratio among affected fields was 0.84\%, suggesting that \ours introduced few irrelevant fields during relation-aware expansion.}
\section{Discussion}

\subsection{Lessons Learned}

\textcolor{black}{\ours is designed for iterative chart refinement rather than one-shot chart generation. It is useful when users know their high-level editing goals but are uncertain about which related parameters may also be affected. By exposing candidate schemes, target fields, and affected fields, \ours turns vague language instructions into inspectable editing spaces and keeps users in control of final design decisions. The detailed comparison with existing natural language-driven chart editing tools is provided in supplemental material.}

\subsection{Generalization and Scalability}

% RAGE-Vis 的可扩展性主要来自其对参数化、可重渲染中间表示的依赖，而不是某一特定图表类型。虽然当前实现主要面向 D3 图表，但若能为 Vega-Lite、ECharts 或 SVG 等表示建立类似的参数抽取与更新机制，RAGE-Vis 的三阶段工作流也有望迁移到更广泛的可视化环境中。此外，关系感知编辑也可作为结构化视觉内容编辑的一种通用思路，扩展到信息图、可视化报告或界面编辑等场景。在这些场景中，局部元素调整同样可能影响文本标注、叙事层级、布局结构或交互反馈，因此也需要识别和呈现相关元素之间的潜在影响，并支持用户进一步调整相关参数。

%The scalability of \ours mainly comes from its reliance on a parameterized and re-renderable intermediate representation, rather than on a specific chart type. Although the current implementation mainly targets D3 charts, if similar parameter extraction and update mechanisms can be established for representations such as Vega-Lite, ECharts, or SVG, the three-stage workflow of \ours may also be transferred to broader visualization environments. In addition, relation-aware editing can serve as a general approach for structured visual content editing, extending to scenarios such as infographic editing, visual report editing, and interface editing. In these scenarios, local element adjustments may also affect text annotations, narrative hierarchy, layout structures, or interaction feedback, and therefore require identifying and presenting potential effects among related elements while supporting users in further adjusting related parameters.
The scalability of \ours mainly depends on whether an input chart can be converted into a parameterized and re-renderable representation. Although our current implementation targets D3 charts, the workflow could be transferred to Vega-Lite, ECharts, SVG, or other structured visualization representations if similar parameter extraction and update mechanisms are available. More broadly, relation-aware editing may also apply to infographics, visual reports, and interface editing, where local changes can affect annotations, layout, narrative hierarchy, or interaction feedback.

\subsection{Limitations and Future Work}

\ours still has several limitations. First, performance bottlenecks stem from chart representation extraction and LLM latency, while image-based reconstruction lacks pixel-level accuracy and complex tasks can degrade interaction responsiveness. Future work will explore robust extraction methods, lightweight model calls, and caching mechanisms to enhance robustness and response efficiency.

Second, the relation scope and quality control are constrained. The three predefined relation types omit complex dependencies like cross-view consistency and high-level design constraints. Lacking automated quality-checking, the system still relies on manual refinement. We plan to integrate visualization design rules and conflict-resolution strategies to provide clearer decision support.

\textcolor{black}{Third, model dependency on a proprietary LLM poses risks for reproducibility. While our training-free workflow generalizes to other structured-output models, weaker models may yield incomplete intents or inconsistent options. Future work will evaluate sensitivity across proprietary and open-source models and implement validation or fallback mechanisms for failure cases.}
\section{Conclusion}

% 本文提出了 RAGE-Vis，一个面向自然语言图表编辑的关系感知生成式界面。针对复合编辑请求、欠指定目标和跨组件一致性维护等挑战，RAGE-Vis 将自然语言请求解析为结构化子意图，并生成由候选方案、目标字段和受影响字段组成的 Visual Panels。通过结合视觉编码关系、结构关系和表达一致性关系，这些面板帮助用户将高层或模糊请求转化为可比较、可调整的编辑空间，使用户能够检查潜在跨组件影响并进一步调整相关字段。案例分析和用户研究表明，RAGE-Vis 能够有效支持高层或欠指定自然语言请求的交互式细化，帮助用户探索候选方案、发现潜在相关字段，并以可控的跨组件联动方式维护图表编辑结果的一致性。
%In this paper, we present \ours, a relation-aware generative editing interface for natural language-based chart editing. To address the challenges of compound editing requests, underspecified goals, and cross-component consistency maintenance, \ours parses natural language requests into structured sub-intents and generates Visual Panels composed of candidate schemes, target fields, and affected fields. By incorporating visual encoding relations, structural relations, and expressive consistency relations, these panels help users transform high-level or vague requests into comparable and adjustable editing spaces, enabling them to inspect potential cross-component effects and adjust related fields. Case studies and user study results show that \ours effectively supports the interactive refinement of high-level or underspecified natural language requests, helps users explore candidate schemes and discover potentially related fields, and maintains the consistency of chart editing results through controllable cross-component coordination.
In this paper, we present \ours, a relation-aware generative editing interface for natural language-based chart editing. \ours parses composite and underspecified requests into structured sub-intents and generates Visual Panels containing candidate schemes, target fields, and affected fields. By incorporating visual encoding, structural, and expressive consistency relations, the system helps users inspect cross-component effects and coordinate related edits. Case studies and a user study show that \ours supports iterative refinement, alternative exploration, related-field discovery, and consistency maintenance in complex chart editing tasks.

% \section*{Supplemental Materials}

\acknowledgments{
We would like to thank the domain experts and anonymous reviewers for their constructive comments. This paper is partially supported by National Natural Science Foundation of China (NO. U23A20313, 62372471).
}

\vspace{1mm}
\noindent
\textbf{CRediT authorship contribution statement}. Ziyao Kang: Software, Data curation, Writing – original draft (equal). Yiping Sun: Investigation, Writing – original draft (equal). Linxuan Tian: Formal analysis, Writing – original draft (equal). Henghuan Qu: Validation. Wei Zeng: Visualization, Methodology(supporting), Writing – review \& editing. Jiazhi Xia: Conceptualization, Methodology (lead), Writing – review \& editing, Funding acquisition, Project administration, Resources, Supervision.

%\acknowledgments{
%The authors wish to thank A, B, and C. This work was supported in part by
%a grant from XYZ.}

%\bibliographystyle{abbrv}
\bibliographystyle{abbrv-doi}

\bibliography{reference}
\end{document}